\documentclass[%
reprint,
amsmath,amssymb,
aps,
]{revtex4-2}

\usepackage{graphicx}
\usepackage{dcolumn}
\usepackage{blindtext}
\usepackage{bm}

\begin{document}

\preprint{asd}

\title{Impact of electron-phonon coupling on electron transport through T-shaped arrangements of quantum dots in the Kondo regime}



\author{Patryk Florków}
\author{Stanisław Lipiński}
\affiliation{Department of Theory of Nanostructures, Institute of Molecular Physics, Polish Academy of Sciences, M. Smoluchowskiego 17, 60-179 Poznań, Poland}
\email[corresponding authors: ]{ patryk.florkow@ifmpan.poznan.pl, \\stanislaw.lipinski@ifmpan.poznan.pl}




\begin{abstract}
We calculate the conductance through strongly correlated T-shaped molecular or quantum dot systems under the influence of phonons.
The system is modelled by the extended Anderson-Holstein Hamiltonian. The finite–U mean-field slave boson approach is used to study 
many-body effects. Phonons influence both interference and correlations. Depending on the dot unperturbed energy and the strength
of electron-phonon interaction the system is occupied by a different number of electrons which effectively interact with each other
repulsively or attractively. This leads together with the interference effects to different spin or charge Fano-Kondo effects.  
\end{abstract}


\maketitle

\section{Introduction}
As the dimension of the mesoscopic system decreases, interactions between electrons become more important and many-body resonances build up. As a consequence, new transport paths are opened. 
The key phenomenon of strong correlations is Kondo effect, which arises from the coherent superposition of cotunneling processes leading to the effective spin flips in consequence of which 
the bound singlet state of the dot spin with the electrons of the leads is formed. This resonance is characterized by SU(2) symmetry. 
In nanoscopic systems SU(2) Kondo effects have been observed in semiconductor-based quantum dots (QDs) \cite{ref1,ref2,ref3,ref4}, in carbon nanotubes \cite{ref5} and in molecular nanostructures \cite{ref6,ref7,ref8,ref9}. 
Besides the spin, also other degrees of freedom, e.g. orbital \cite{ref10} or charge \cite{ref11, ref12} can give rise to Kondo correlations. For systems with higher degeneracy e.g. 
in the case of fourfold spin – orbital degeneracy not only spin, but also orbital pseudo-spin can be screened. Such SU(4) Kondo effect resonances have been observed in vertical QDs \cite{ref10}, 
in capacitively coupled dots \cite{ref13} and in carbon nanotubes \cite{ref14,ref15,ref16,ref17}. 

There is currently also a great interest in the interplay of strong correlations and interference 
in multiply connected geometries e.g. in T-shaped systems, where dot or molecule is side-coupled to a quantum wire \cite{ref18, ref19, ref20, ref21, ref22, ref23, ref24, ref25, ref26}. 
Sidewall chemical functionalization of molecular wires is already a well-established branch of research. The attached objects act as scatterers for electron transmission through the quantum wire 
and allow to tune its transport properties. In T-shaped systems, the interference of different conduction paths can lead to Fano antiresonance manifesting as a dip in the linear conductance \cite{ref23, ref24, ref27, ref28}. 
There are also reports on the T-shaped carbon nanotube structures \cite{ref29, ref30} and similar carbon devices engineered by attaching C60 buckyballs onto the sidewall of a single-walled carbon nanotube (carbon nanobud \cite{ref31}). 
Many experiments showed that Kondo and Fano resonances can occur simultaneously \cite{ref32, ref33}. 

Recently there is also an increasing interest in nanoelectromechanical systems (NEMS) integrating electrical and mechanical 
functionality \cite{ref34,ref35,ref36,ref37,ref38}. Nanoelectromechanical systems utilizing localized mechanical vibrations have found applications in ultrafast sensors, actuators and signal processing components. 
Of special interest are molecular systems, because molecules due to their softness easy deform during tunneling processes giving rise to excitation of local phonon modes. 
The polaronic transport through molecular systems has been recently studied in a number of papers e.g. \cite{ref39,ref40,ref41,ref42,ref43,ref44}. 
Due to participation of localized phonon in single electron tunneling the phonon side bands appear in the spectral function of the dot. 
Interesting, similar effects have been also observed in the rigid structures of semiconductor quantum dots embedded in a freestanding GaAs/AlGaAs membrane \cite{ref44,ref45,ref46,ref47,ref48}. 
It has been shown that morphology manipulation of semiconductor QDs such as size shape, strain distribution or inhomogenities can influence coupling strength of e-ph interaction \cite{ref49}. 
The phononic effects exist not only in the sequential tunneling, but also in Kondo regime where the vibrational sidebands have been also observed \cite{ref50, new17, new18, ref45, new19, ref51, new20}. 
The interplay of electron-phonon coupling and Kondo effect has been also studied theoretically \cite{ref51,ref52,ref53,ref54,ref55,ref56}.
In the present paper we analyze impact of electron - phonon coupling on both strong correlations and interference. 
We perform a discussion for quantum dots arranged in single (TQD) or double (DTQD) T- shaped geometries (Fig. \ref{fig1}). Due to the quantum confinement, there may be also confined phonon located in a single QD or molecule. 
Such a phonon interacts only with the electrons in the same QD. 
In the following considerations it is assumed that local phonons couple either to the open dots (OQDs) directly connected to the leads or to the interacting (IQDs) side attached dots. 
The former type of coupling although mainly influences interference conditions, it also affects the correlations. In the second case, only the correlations are modified. 
For phonons coupled to OQDs roughly exponential suppression of transmission and the occurrence of the satellite Fano-Kondo dips is observed. 
These effects manifest only very weakly in the transmission through the open dot in the case when phonons couple to IQDs. However, they reflect clearly in the density of states (DOS) of IQDs, 
but this is difficult to detect in transport experiments. The single T-shaped device decoupled from phonons is characterized by SU(2) symmetry and electron occupation ranges from zero to two, 
and double T-shaped device unperturbed by phonons has SU(4) symmetry with possible electron occupations from zero to four. 
Electron-phonon (e-ph) Holstein coupling which we discuss, lowers the energy of doubly occupied orbitals relative to the single occupied or empty. 
In consequence the regions of occurrence of an even number of electrons in the system narrow down or completely disappear with the increase of the strength of e-ph interaction. 
Fluctuating spin doublets interfering with the wave propagating through direct path give rise through cotunneling processes to the spin Fano-Kondo effect. 
For strong e-ph coupling the charge ordered ground state of the system appears and for special ranges of gate voltages and the corresponding value of the e-ph coupling the charge Fano-Kondo effect occurs. 

\section{Model and formalism}
\begin{figure}
\centering
\includegraphics[width=4.5cm]{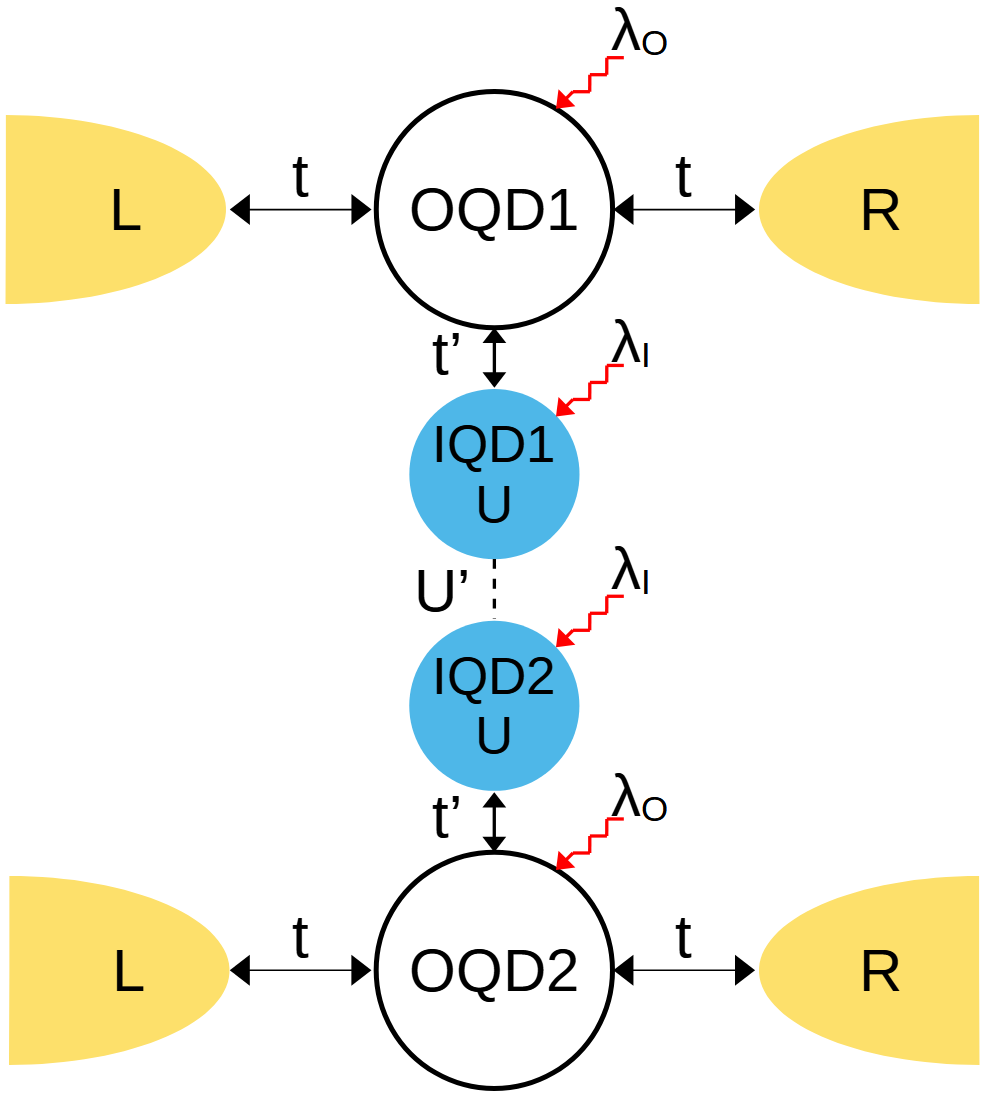}
\caption{Schematic of capacitively coupled side attached quantum dots IQDs (DTQD) with electron-phonon coupling. 
The electrodes (L, R) are directly attached to the open quantum dots (vanishing interactions) OQD1(2). 
Phonons are coupled either to open $(\lambda_{O}\neq 0)$ or to interacting dots $(\lambda_{I}\neq 0)$.}
\label{fig1}
\end{figure}

We consider two types of T-shaped structures. The first is a single T-shaped system, in which one of the dots (noninteracting dot - OQD) is coupled directly to the leads and the second dot (interacting - IQD) is coupled indirectly, 
through the open dot. The scheme presented on Fig. \ref{fig1} shows two capacitively coupled TQD systems i.e. DTQD.
It is worth noting that orbitally degenerate TQD system with equal intra and interorbital interactions is equivalent to DTQD we consider. 
This interpretation of DTQD systems is more adequate for molecular systems. Both side coupled dot systems, single TQD and double DTQD with Kondo resonances of interacting dots have already been analyzed \cite{ref22, ref24, ref57,ref58,ref59,ref60}, 
but here we generalize these considerations focusing on the role of phonons in electron transport through these structures.
Discussing the electron–phonon coupling in the introduced systems we consider three special cases: 1) local phonon modes are coupled solely to the open dots (l=1), 
2) local phonon modes are coupled only to the interacting dots (l=2) or 3) single local phonon mode is equally coupled to both interacting dots (l=3). 
The corresponding DTQD Hamiltonians representing the three mentioned e-ph types of coupling are written below:
\begin{eqnarray}
\mathcal{H}^{(l)} = \mathcal{H}_{A}^{DTQD} + \mathcal{H}_{ph}^{DTQD(l)} + \mathcal{H}_{e-ph}^{DTQD(l)},
\end{eqnarray}
where $\mathcal{H}_{A}^{DTQD}$ is the double dot or double orbital Anderson Hamiltonian for T-shaped geometry which is written as: 
\begin{eqnarray}
\mathcal{H}_{A}^{DTQD} =  \sum_{k\alpha j\sigma} \epsilon_{k\alpha j}c_{k\alpha j\sigma}^{\dagger}c_{k\alpha j\sigma} 
+ \sum_{j\sigma}E_{O}d_{j\sigma}^{\dagger}d_{j\sigma} \nonumber \\ + \sum_{j\sigma} E_{f}f_{j\sigma}^{\dagger}f_{j\sigma}
+ \sum_{j}U n_{j\uparrow}n_{j\downarrow} + \sum_{\sigma \sigma '} U' n_{1\sigma}n_{2\sigma '} \nonumber\\
+ \sum_{k\alpha j\sigma}t(c_{k\alpha j\sigma}^{\dagger}d_{j\sigma} + h.c.) + \sum_{k\alpha j\sigma}t'(d_{j\sigma}^{\dagger}f_{j\sigma} + h.c.),
\end{eqnarray}
where the first term describes electrons in the electrodes and the next two terms represent electrons residing on the open ($d_{j\sigma}$) and interacting ($f_{j\sigma}$) dots respectively, 
j enumerates the upper (j=1) and the lower (j=2) TQD subsystems visualized on Fig. \ref{fig1} and $\alpha$ numbers left or right electrode.
The terms parametrized by U, U’ describe intra and interdot Coulomb interactions, with $n_{j\sigma} = f^{\dagger}_{j\sigma}f_{j\sigma}$ denoting occupation operators of IQDs 
and the last two terms stand for tunneling between the electrodes and open dots and between the dots. 
In case of no e-ph coupling intra and interdot interactions are assumed to be equal (U=U'). $\mathcal{H}_{ph}$ is phonon Hamiltonian and $\mathcal{H}_{e-ph}$ is Holstein electron-phonon coupling term \cite{ref61}.
\begin{align}
\mathcal{H}_{ph}^{(1)} = \mathcal{H}_{ph}^{(2)} = \omega_{0}\sum_{j}b_{j}^{\dagger}b_{j} \nonumber\\
\mathcal{H}_{ph}^{(3)} = \omega_{0}b^{\dagger}b 
\end{align}
where $\omega_{0}$ is the frequency and $b_{j}$ annihilation operator of the localized phonon mode (we set $\hbar = k_{B} = |e| = 1$).
\begin{align}
\mathcal{H}_{e-ph}^{(1)} = \sum_{j\sigma}\lambda_{O}(b_{j}^{\dagger} + b_{j})d_{j\sigma}^{\dagger}d_{j\sigma}, \nonumber\\
\mathcal{H}_{e-ph}^{(2)} = \sum_{j\sigma}\lambda_{I}(b_{j}^{\dagger} + b_{j})f_{j\sigma}^{\dagger}f_{j\sigma}, \nonumber\\
\mathcal{H}_{e-ph}^{(3)} = \lambda_{I} (b^{\dagger} + b) \sum_{j\sigma}f_{j\sigma}^{\dagger}f_{j\sigma}
\end{align}
Dot Hamiltonian for TQD system is
\begin{eqnarray}
\mathcal{H}^{TQD(l)}=\mathcal{H}_{A}^{TQD} + \mathcal{H}_{ph}^{TQD(l)} + \mathcal{H}_{e-ph}^{TQD(l)} 
\end{eqnarray}
We do not write explicit forms of $\mathcal{H}_{A}^{TQD}$ and $\mathcal{H}_{ph}^{TQD(l)}$, $\mathcal{H}_{e-ph}^{TQD(l)}$ (l=1,2) since they only differ from (2,3,4) by a lack of summation over the phonon modes and dots. 
In this case only a single phonon couples to electrons. The term describing the interaction between the dots parameterized by U’ does not appear either. 

Following Lang and Firsov \cite{ref62, ref63} the electron-phonon couplings in DTQD can be eliminated by canonical transformations:
\begin{eqnarray}
\mathcal{H}^{DTQD(l)}=e^{iS(l)}\mathcal{H}^{DTQD(l)}e^{-iS(l)},
\end{eqnarray}
with
\begin{eqnarray}
S(1)=\frac{\lambda_{O}}{\omega_{0}}\sum_{j\sigma}(b_{j}^{\dagger}-b_{j})d_{j\sigma}^{\dagger}d_{j\sigma}, \nonumber\\
S(2)=\frac{\lambda_{I}}{\omega_{0}}\sum_{j\sigma}(b_{j}^{\dagger}-b_{j})f_{j\sigma}^{\dagger}f_{j\sigma}, \nonumber\\
S(3)=\frac{\lambda_{I}}{\omega_{0}}(b^{\dagger}-b)\sum_{j\sigma}f_{j\sigma}^{\dagger}f_{j\sigma}
\end{eqnarray}
The new fermion (polaron) operators are $\widetilde{f}_{j\sigma}=f_{j\sigma}X_{j}$ and 
$\widetilde{f}_{j\sigma}^{\dagger}=f_{j\sigma}^{\dagger}X_{j}^{\dagger}$ with $X_{j}^{(1)}=exp[-\frac{\lambda_{O}}{\omega_{0}}(b_{j}^{\dagger}-b_{j})]$ (l=1) 
and similarly for the coupling with interacting dots $\widetilde{d}_{j\sigma}=d_{j\sigma}X_{j}$ and $\widetilde{d}_{j\sigma}^{\dagger}=d_{j\sigma}^{\dagger}X_{j}^{\dagger}$ with 
$X_{j}^{(2)}=exp[-\frac{\lambda_{I}}{\omega_{0}}(b_{j}^{\dagger}-b_{j})]$ for l=2 and $X_{j}^{(3)}=exp[-\frac{\lambda_{I}}{\omega_{0}}(b^{\dagger}-b)]$ for $l=3$.

The diagonalization \cite{ref6} is exact if $t=t’=0$ or $\lambda_{O}=\lambda_{I}=\infty$. 
This transformation shifts the dots to the new equilibrium positions and in general changes the phonon vacuum. 
In the following we restrict in the expansion of $\mathcal{H}$ up to the terms $\lambda^{2}$. 
In this sense, the results for stronger coupling should be viewed with caution, 
treating them only as a quantitative, preliminary insight into the problem. 
Higher commutator approximations improve accuracy of canonical transformation, 
but introduce numerical difficulties \cite{new1}. 
Very strong coupling regime is usually described starting from infinite coupling solution and then performing perturbation expansion in terms of $1/\lambda$ \cite{new2}.
Analogous unitary transformations decoupling the entanglement of electrons and phonons in TQD systems have the same form, but again without summing over index $j$. 
The DTQD Hamiltonians are transformed into $\widetilde{\mathcal{H}}^{DTQD(l)}=\widetilde{\mathcal{H}}_{A}^{DTQD}+\mathcal{H}_{ph}(l)$. $\widetilde{\mathcal{H}}_{A}^{DTQD}$ has the same form as $\mathcal{H}_{A}^{DTQD}(2)$, 
but with old fermion operators replaced by new operators $d_{j\sigma} \rightarrow \widetilde{d}_{j\sigma}$ for $l=1$ or $f_{j\sigma} \rightarrow \widetilde{f}_{j\sigma}$ for $l=2, 3$ and the use of renormalized parameters. 
For $l=2, 3$ the parameters $E_{f}$ and $U$ are shifted due to e-ph interaction by a renormalization constant $\lambda_{I}^{2}/\omega_{0}$, $\widetilde{E}_{f}=E_{f}-\lambda_{I}^{2}/\omega_{0}$, 
$\widetilde{U}=U-2\lambda_{I}^{2}/\omega_{0}$ (l=2) and $\widetilde{E}_{f}=E_{f}-\lambda_{I}^{2}/\omega_{0}$, $\widetilde{U}=U-2\lambda_{I}^{2}/\omega_{0}$, $\widetilde{U'}=U'-2\lambda_{I}^{2}/\omega_{0}$ (l=3). 
As it is seen Holstein coupling lowers the energy of doubly occupied orbitals relative to single occupied or empty. For l=1 e-ph interaction effectively shifts $E_{O}$, $\widetilde{E}_{O}=E_{O}-\lambda_{O}^{2}/\omega_{O}$ 
and attractive phonon induced interaction $-2(\lambda_{O}^{2}/\omega_{O})n_{f\uparrow}n_{f\downarrow}$ appears. 
For l=2, 3 hopping term between the open and interacting dot is also renormalized by a factor $X$, which describes the effect of the phonon cloud accompanying the hopping process $\widetilde{t'}=t'X$. 
For l=1 both hopping integrals are renormalized $\widetilde{t}=tX_{j}$ and $\widetilde{t'}=t'X_{j}$. 
Assumption of the relaxation time of phonons to be much shorter than the time of electron transport through the dot (antiadiabatic limit) allows to consider the phonon subsystem as being approximately in thermal equilibrium. 
The phonon operator $X$ can be then replaced with its expectation value \cite{ref63} $\langle X\rangle=exp[-(\lambda^{2}/\omega_{0})(N_{ph}+1/2)]$ with $\lambda=\{\lambda_{I},\lambda_{O}\}$, depending on the analyzed case and 
$N_{ph}$ is given by the Bose-Einstein distribution. In this approach the electron and phonon dynamics become decoupled. This widely used in literature approximation (e.g. \cite{ref49, ref54, ref55, ref64,ref65,ref66}) 
predicts of exponential suppression of the tunneling amplitudes (Franck-Condon type suppression). Using the form of new fermion operators obtained in Lang-Firsov transformation 
it is easy to show that the dot electron Green’s function can be decoupled as (e.g. \cite{ref67, ref68}):
\begin{eqnarray}
G_{dj\sigma}(t)=-i\theta(t)\langle[\widetilde{d}_{j\sigma}(t), \widetilde{d}_{j\sigma}^{\dagger} (0)]\rangle = \nonumber\\
-i\theta(t)\langle[ e^{i(\mathcal{H}_{e}+\mathcal{H}_{ph})t} \widetilde{d}_{j\sigma} e^{-i(\mathcal{H}_{e} + \mathcal{H}_{ph})t} \widetilde{d}_{j\sigma}^{\dagger} ]\rangle = \nonumber\\
-i\theta(t) \{ \langle d_{j\sigma}'(t) d_{j\sigma}^{\dagger} \rangle_{e}\langle X(t) X^{\dagger} (0)\rangle_{ph} + \nonumber\\
\langle d_{j\sigma}^{\dagger \prime}(0) d_{j\sigma}(t) \rangle_{el} \langle X^{\dagger}(0) X(t)\rangle_{ph} \},
\end{eqnarray}
where $\widetilde{d}_{j\sigma}(t)=e^{i\mathcal{H}_{e}t}\widetilde{d}_{j\sigma}e^{-i\mathcal{H}_{e}t}$, $X(t)=e^{i\mathcal{H}_{ph}t}X e^{-i\mathcal{H}_{ph}t}$, and $\mathcal{H}_{e}=\mathcal{H}_{A}^{DTQD}$
or $\mathcal{H}_{ph}=\mathcal{H}_{A}^{DTQD}$ respectively. The renormalization factor due to e-ph interaction is evaluated as \cite{ref63}:
\[
\langle X(t) X(0)^{\dagger}\rangle_{ph}=exp(-\theta (t)),
\]
where
\begin{align}
\theta (t)=(\frac{\lambda}{\omega_{0}})^{2}[N_{ph}(1-e^{i\omega_{0}t})+(N_{ph}+1)(1-e^{-i\omega_{0}t})]
\end{align}
For the zero-temperature case we are discussing a situation where $\langle X(t)X^{\dagger}(0)\rangle$ is reduced to
\begin{eqnarray}
\langle X(t)X^{\dagger}(0)\rangle=\sum_{n=0}^{\infty}L_{n}e^{in\omega_{0}t},
\end{eqnarray}
where
\begin{eqnarray}
L_{n}=\frac{1}{n!}(\frac{\lambda}{\omega_{0}})^{2n}e^{-(\frac{\lambda}{\omega_{0}})^{2}}  
\end{eqnarray}
and this approximate formula will be used by us in the following, because we are interested only in the low temperature transport. The Fourier transforms of the retarded Green's functions for the dots, 
to which the phonons are attached is then given by
\begin{align}
\textbf{G}_{dj\sigma}^{R}(\omega)=\sum_{n=-\infty}^{\infty}L_{n}[1-f(\omega-n\omega_{0})\widetilde{G}_{dj\sigma}(\omega-n\omega_{0}) \nonumber\\
+ f(\omega+n\omega_{0})\widetilde{G}_{dj\sigma}(\omega+n\omega_{0})],
\end{align}
where $f(\omega)$ is a Fermi distribution function and the retarded dressed Green's functions $\textbf{G}_{dj\sigma}^{R}(\omega)$ are the functions corresponding 
to Hamiltonian (6) for  l=2, 3. Analogous expression to (11) holds for $\textbf{G}_{fj\sigma}^{R}(\omega)$ (Hamiltonian (6) l=2, 3). 
To find the dressed Green’s functions and consequently discuss correlation effects we use finite U slave boson mean field approximation (SBMFA) 
of Kotliar and Ruckenstein \cite{ref69, ref70}, which we apply to the effective polaron Hamiltonian of DTQD (6) or analogous Hamiltonian for TQD. 
For the latter case a set of auxiliary bosons $e$, $p_{\sigma}$ and $d$ projecting onto empty, single occupied and doubly occupied states of interacting QD 
are introduced. In DTQD system apart from $e$, $p$, $d$ operators we also introduce $t$ and $f$ SB operators representing triple and quadruple dot fillings respectively. 
The single occupation projectors $p_{j\sigma}$ in this case are additionally labeled by orbital or interacting dot index j. 
Similar notation applies for triple occupancy boson $f_{j\sigma}$, but this time index j indicates interacting dot or orbital, which is not fully occupied 
(occupation of a hole). Six $\{d_{i},d_{\sigma\sigma'}\}$ operators project onto $(\uparrow\downarrow, 0)$ and $(0,\uparrow\downarrow)$ for $(d_{j=1,2})$
and $(\uparrow,\uparrow),(\downarrow,\downarrow),(\uparrow,\downarrow),(\downarrow,\uparrow)$ for $(d_{\sigma \sigma'})$ \cite{ref71}. To eleminate unphysical states 
we introduce constraints: completeness relation for the slave boson operators and the condition for the correspondence between fermions and bosons. 
These restrictions can be enforced by introducing Lagrange multipliers $(\Delta',\Delta_{j\sigma})$ and supplementing the effective slave boson Hamiltonian by 
corresponding terms. For brevity we write only SBMFA Hamiltonian of DTQD for l=2.
\begin{align}
\mathcal{H}_{l=2}^{K-R}=\sum_{k\alpha j\sigma}\epsilon_{k\alpha j\sigma}c_{k\alpha j\sigma}^{\dagger}c_{k\alpha j\sigma} + \sum_{j\sigma}E_{O}d_{j\sigma}^{\dagger}d_{j\sigma} \nonumber\\
+ \sum_{j\sigma}\widetilde{E}_{jf}\textit{f}_{j\sigma}^{\dagger}\textit{f}_{i\sigma}
+ \widetilde{U}\sum_{j=1}^{2}d_{j}^{\dagger}d_{j} + U'\sum_{\sigma\sigma'}d_{\sigma\sigma'}^{\dagger}d_{\sigma\sigma'} \nonumber\\ 
+ (\widetilde{U}+2U')\sum_{j\sigma}t_{j\sigma}^{\dagger}t_{j\sigma} + (2\widetilde{U}+4U')f^{\dagger}f \nonumber\\
+ \sum_{j\sigma}\Delta_{j\sigma}(n_{fj\sigma}-Q_{j\sigma})+\Delta'(I-1) \nonumber\\
+ t\sum_{k\alpha j\sigma}(c_{k\alpha j\sigma}^{\dagger}d_{j\sigma}+h.c.)+t'\sum_{j\sigma}(d_{j\sigma}^{\dagger}z_{j\sigma}\textit{f}_{j\sigma}+h.c.)
\end{align}
where pseudofermion operators $\textit{f}_{j\sigma}$ are defined by $\widetilde{\textit{f}}_{j\sigma}=\textit{f}_{j\sigma}z_{j\sigma}$, and 
$Q_{j\sigma}=p_{j\sigma}^{\dagger}p_{j\sigma}+d_{j\sigma}^{\dagger}d_{j\sigma}+d_{\sigma\sigma}^{\dagger}d_{\sigma\sigma}+d_{\sigma\overline{\sigma}}^{\dagger}d_{\sigma\overline{\sigma}}
+t_{\overline{j}\sigma}^{\dagger}t_{\overline{j}\sigma}+t_{\overline{j\sigma}}^{\dagger}t_{\overline{j\sigma}}+\textit{f}^{\dagger}\textit{f}$, 
$I=e^{\dagger}e+\textit{f}^{\dagger}\textit{f}+\sum_{j\sigma}(p_{j\sigma}^{\dagger}p_{j\sigma}+t_{j\sigma}^{\dagger}t_{j\sigma})+\sum_{j}d_{j}^{\dagger}d_{j}+\sum_{\sigma\sigma'}d_{\sigma\sigma'}^{\dagger}d_{\sigma\sigma'}$ and
$z_{j\sigma}=(e^{\dagger}p_{j\sigma}+p_{j\sigma}+p_{j\overline{\sigma}}^{\dagger}d_{j}+p_{\overline{j\sigma}}^{\dagger}(\delta_{j,1}d_{\sigma\overline{\sigma}}+\delta_{j,2}d_{\overline{\sigma}\sigma})
+p_{\overline{j}\sigma}^{\dagger}d_{\sigma\sigma}+d_{\overline{j}}^{\dagger}t_{j\sigma}+d_{\overline{\sigma\sigma}}^{\dagger}t_{\overline{j\sigma}}+(\delta_{j,2}d_{\sigma\overline{\sigma}}^{\dagger}+
\delta_{j,1}d_{\overline{\sigma}\sigma}^{\dagger})t_{\overline{j}\sigma}+t_{j\overline{\sigma}}\textit{f})/\sqrt{Q_{j\sigma}(1-Q_{j\sigma})}$. $z_{j\sigma}$ renormalize interdot hoppings and dot-lead hybridization. The stable mean field solutions are found
from the minimum of the free energy with respect to the mean values of boson operators and Lagrange multipliers. For the high symmetry coupling cases discussed, where two Kondo dots play identical role the number of 
independent bosons is reduced to six $(e,p,d,d',t,\textit{f})$ and only two Lagrange coefficients are enough $(\Delta, \Delta')$. In SBMFA procedure the problem of strong interactions is formally reduced to the effective free
electron model with renormalized hopping integrals and dot energies. SBMFA best describes systems close to the unitary Kondo limit, but due to its simplicity, this approach is also often used in analysis of linear conductance 
of systems with weakly broken symmetry giving results in a reasonable agreement with experiment and with numerical renormalization group calculations \cite{ref63}. Mean field approximation best works at low temperatures, where 
it is justified to neglect fluctuations of the boson fields. We restrict our analysis to equilibrium, so inelastic transport produced by e-ph interaction can be neglected.
We consider conductances through the upper TQD subsystem ($j=1$) and through the lower subsystem ($j=2$). They are separately experimentally accessible.
According to the derivation based on the nonequilibrium Green function formalism \cite{new3} linear conductances of the wires with embeded open dots are given by Landauer type formula:

\begin{eqnarray}
\mathcal{G}_{j\sigma}=\frac{e^{2}}{\hbar}\frac{\Gamma}{2}\int_{-\infty}^{\infty}d\omega(-\frac{\partial f(\omega)}{\partial \omega})\mathcal{I}m\textbf{G}_{dj\sigma}^{R}(\omega),
\end{eqnarray}
where $\Gamma$ is the coupling strength to the electrodes (for the rectangular density of states of electrodes $1/2D$ for $|E|<D$, $\Gamma=\pi t^{2}/D$). 
For the case when phonons are coupled to the open dot $\Gamma$ should be replace by $\widetilde{\Gamma}=\pi \widetilde{t}^{2}/D$, $\widetilde{t}^{2}=t^{2}exp(-\lambda/\omega_{0})^{2}$, 
$\textbf{G}_{dj\sigma}$ denotes the Green's function of OQDj, which according to (11) can be approximately expressed as 

\begin{align}
\textbf{G}_{dj\sigma}^{R}(\omega)= \nonumber\\ \sum_{n=-\infty}^{\infty}L_{n}\{(1 - f(\omega-n\omega_{0})) 
\frac{1}{\omega-n\omega_{0}-\widetilde{E}_{0}+i\widetilde{\Gamma}} \nonumber\\
+ f(\omega+n\omega_{0}) \frac{1}{\omega+n\omega_{0}-\widetilde{E}_{0}+i\widetilde{\Gamma}} \nonumber\\
(1+\frac{|\widetilde{t}'|^{2}}{\omega-\widetilde{E}_{0}+i\widetilde{\Gamma}}) \textbf{G}_{fj\sigma}^{(O)R}(\omega)\}
\end{align}
with
\begin{eqnarray*}
\textbf{G}_{fj\sigma}^{(O)R}=\frac{1}{\omega-\widetilde{E}_{d}-\Delta_{j\sigma}(l=1)-\frac{\widetilde{t}^{2}z_{j\sigma}^{2}}{\omega-\widetilde{E}_{0}+i\widetilde{\Gamma}}}
\end{eqnarray*}
for l=1 and
\begin{eqnarray*}
\textbf{G}_{dj\sigma}^{R}(\omega)=\frac{1}{\omega-\widetilde{E}_{0}+i\widetilde{\Gamma}}(1+\frac{|\widetilde{t}'|^{2}}{\omega-\widetilde{E}_{0}+i\widetilde{\Gamma}}) \nonumber\\ 
\{\sum_{n=-\infty}^{\infty}L_{n}(1-f(\omega-n\omega_{0}))\textbf{G}_{fj\sigma}^{(O)R}(\omega-n\omega_{0}) \nonumber\\
+f(\omega+n\omega_{0})\textbf{G}_{fj\sigma}^{(O)R}(\omega+n\omega_{0})\},
\end{eqnarray*}
for l=2, 3 with
\begin{eqnarray}
\textbf{G}_{fj\sigma}^{(O)R}=\frac{1}{\omega-\widetilde{E}_{d}-\Delta_{j\sigma}(l)-\frac{\widetilde{t}^{2}z_{j\sigma}^{2}}{\omega-\widetilde{E}_{0}+i\widetilde{\Gamma}}}
\end{eqnarray}

\section{Results and discussion}
Before presenting numerical results let us look at the energy scales we are going to discuss. 
Typical values of the bandwidths of the metallic organic wires are several hundreds of $meV$ \cite{new4, new5} Charging energy parametrized by $U$ can be inferred from the size of the Coulomb diamonds. 
It increases with the decrease of the size of QD. Typically for molecular and semiconducting QDs it ranges from several to tens of $meV$ \cite{new6, new7, new8, new9}. 
Coupling between the QD and reservoirs $\Gamma$ can be estimated from the width of Coulomb peaks and in the interesting us weak-coupling regime ranges from hundreds $\mu eV$ up to several $meV$ \cite{new7, ref14}.
Phonon energies of molecular systems range from $\mu eV$ up to $meV$ \cite{new10, new11}. 
The experimental values for the electron-phonon coupling strength depend on the specific setup. All the ranges: weak, intermediate and strong are accessible. As an example, 
let us only quote data for carbon systems. The experimental results for suspended carbon nanotubes QDs showed an average value of strong coupling $\lambda\approx 1.7$ \cite{ref48}. 
For fullerene C60 intermediate coupling is observed $\lambda\approx 0.5$ \cite{ref45} 
and for different C140 samples $\lambda$ ranges between 0.1 and 4 \cite{new12, new13}.
In this section we present numerical calculations illustrating the effect of phonons on interference and electron correlations and we show how it is reflected in conductance. 
Throughout this paper we use relative energy units choosing $D/50$ as the unit, where $D$ is the electron bandwidth of the leads. We assume the following values of parameters: 
the bare Coulomb integrals in the absence of phonons $U=U’=3$, coupling strength to the electrodes in the range $\Gamma\in \langle 0.05, 0.1\rangle$, 
and phonon energy $\omega_{0}=0.5$. The Fermi energy of the leads is taken as $E_{F}=0$. 
All the results have been calculated for the strong e-ph coupling limit $(\lambda_{O}, \lambda_{I} > \Gamma)$ and compared with the cases without phonons $(\lambda_{O}=0, \lambda_{I}=0)$. 
The shape of the bare transmission lines of the considered T-shaped systems is determined by Fano parameter $q = E_{O}/\Gamma$, which can be tuned by gate voltage. 
For $q=0$ interference between the ballistic channel through the open dot and the Kondo resonant channel leads to the symmetric dip structure with vanishing transmission for SU(2) symmetry 
(destructive interference, Fano-Kondo antiresonance) and to half reflection for SU(4). This is a consequence of $\pi/2$ or $\pi/4$ phase shifts for SU(2) or SU(4) symmetries correspondingly. 
A complete reflection for SU(4) symmetry occurs for $q=-1$ (destructive interference) and full transmission for $q=1$ (constructive interference). 
SU(2) T-shaped system exhibits unitary transmission for $q=|\infty|$ and in this limit it is roughly equivalent to the embeded QD \cite{ref58}. 

\subsection{Single T-shaped quantum dot structure}

\begin{figure*}
\centering
\includegraphics[width=16cm]{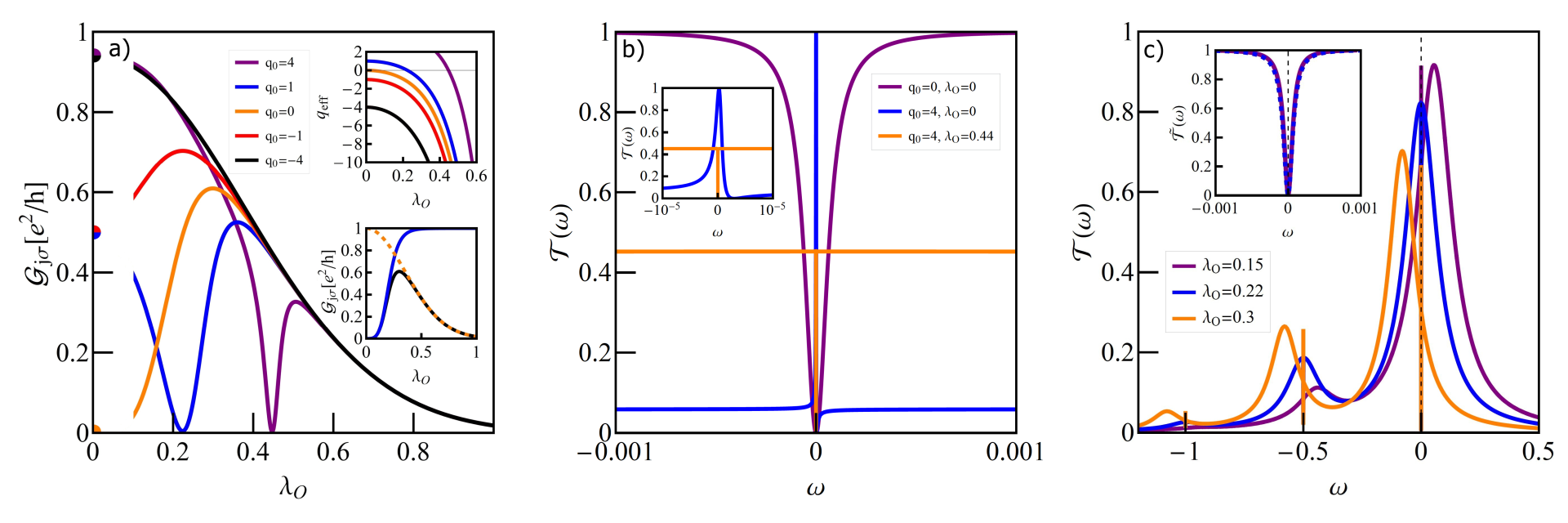}
\caption{(a) Partial conductance of a single T-shaped system TQD with phonon mode coupled to OQD plotted for different values of Fano factor $q_{0}$ $(E_{f}=-1.5)$. 
Inset at the top of Fig. a shows effective Fano factor as a function of $\lambda_{O}$. The bottom inset illustrates the impact of Franck-Condon type suppression. 
The orange–broken line presents F-C factor, black solid line shows dependence of conductance on $\lambda_{O}$ for $q_{0}=0$ resulting from both effects, 
renormalization of $q$ and F-C suppression and blue solid line shows the effect on conductance of only phonon induced renormalization of $q$. 
(b) Selected transmissions of TQD for $\lambda_{O}=0$, $q_{0}=0$, $q_{0}=4$ and for $\lambda_{O}=0.44$, $q_{0}=4$. 
(c) Transmission of TQD in a wide range of energy $\omega$ showing the traces of phonon modes around $n\omega_{0}$. In the inset normalized transmission for $\lambda_{O}=0.22$ around $\omega =\omega_{0}$ 
is imposed on the normalized transmission around $\omega=0$.}
\label{fig2}
\end{figure*}

Fig. \ref{fig2}a illustrates the dependence of conductance on e-ph coupling strength of TQD for the case when local phonon is coupled to the open dot. 
The curves are plotted for different unperturbed Fano parameters. For vanishing coupling $(\lambda_{O}=0)$ interplay of Kondo correlations and interference results in forming Fano-Kondo antiresonance for $q=0$ (Fig. \ref{fig2}b) 
and consequently zero conductance is observed. For $q\neq 0$ Fano-Kondo resonances are asymmetric and dip does not enter the Fermi level and corresponding conductances are finite. 
For $q\rightarrow \infty$ the resonance evolves into Lorentzian peak at $E_{F}$ \cite{new14}. 
It is already partially visible in the transmission for $q=4$, $\lambda_{O}=0$ (Fig. \ref{fig2}b) and then conductance reaches then nearly unitary limit. 
Looking at the transmission for $q=4$ and $\lambda_{O}=0.44$, where effective Fano parameter vanishes, it is seen that the symmetric antiresonance is rebuilt, 
but transmission is suppressed due to e-ph coupling. It can be seen that the coupling with phonons modifies the interference conditions. 
The effective Fano parameters $q_{eff}$, presented in the upper inset of Fig. \ref{fig2}a decrease with increasing coupling. $q_{eff}$ are determined by coupling dependent effective site energy of the open dot 
$\widetilde{E}_{O}$ through polaron shift and phonon dependent hybridization strength $\widetilde{\Gamma}$, which changes according to the Franck-Condon factor (F-C), 
$exp[-(\lambda_{O}/\omega_{0})^{2}]$. Lower inset shows the examples of conductance dependence on $\lambda_{O}$ with the inclusion or neglect of F-C factor. 
The exponentially decreasing line illustrates the pure Franck-Condon suppression (conductance for $q=\infty$). It is seen that for small values of $q$ this suppression starts to play the decisive role for large values of coupling. 
For small values of e-ph coupling linear conductance increases for $q\le 0$, with the increase of $\lambda_{O}$ and decreases for $q>0$. The observed decrease of conductance for strong e-ph coupling is dictated by F-C suppression. 
The zeros of conductance correspond to $q_{eff}=0$. Fano-Kondo resonance is also influenced by phonons through the change of Kondo correlations resulting mainly from the renormalization of interdot hopping $t'$ and 
transmission variations on the open dot induced by the changes of electron-phonon coupling. Fig. \ref{fig2}c shows the examples of transmissions for $q_{0}=1$ for the selected values of $\lambda_{O}$. 
The main peak and the satellites move towards lower energies with the increase of e-ph coupling, what corresponds to the phonon induced shift of $E_{O}$. The height of the main peak decreases according to F-C factor. 
The dips observed for $\omega=0$ for the main peaks (see the inset) and for $\omega=n\omega_{0}$ for the satellites, exhibit Fano shape corresponding to a given $q_{eff}$. For $\lambda_{O}=0.22$, where $q_{eff}=0$ the line is symmetric. 
It is worth to stress that the widths of the dips in the main peak are the same as in the satellites, what proves that both reflect the same phenomena - Kondo resonance on the interacting dot. 
We have also checked that the satellite peaks follow the same temperature dependence as the main Kondo peak. So far we have presented conductance for occupation $n=1$ only. 
Next figures \ref{fig3}-\ref{fig5} refer to TQD system with phonon attached to an interacting dot. 
\begin{figure*}
\centering
\includegraphics[width=12cm]{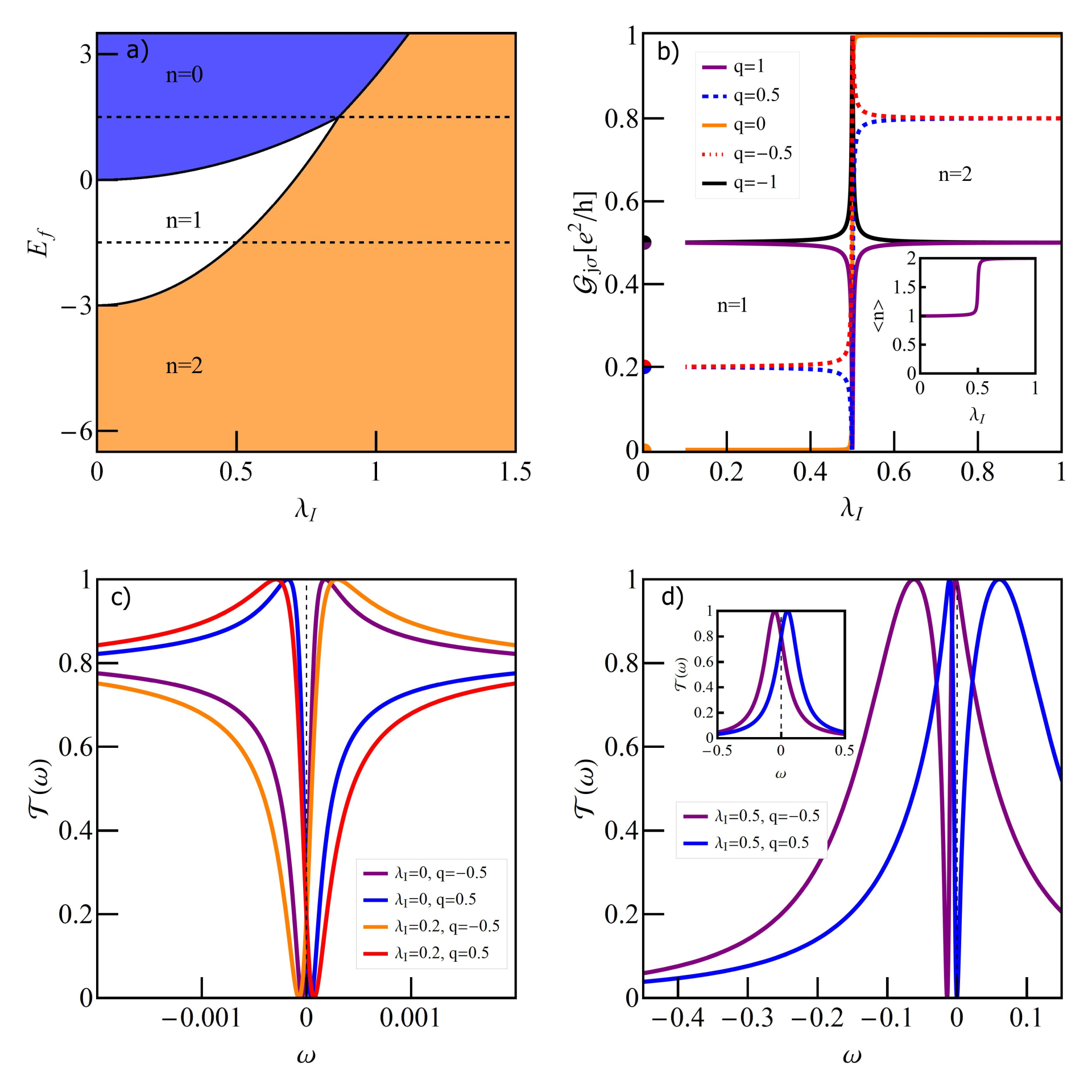}
\caption{(a) Charge stability diagram of TQD with phonon coupled to IQD as a function of gate voltage and e-ph coupling $\lambda_{I}$. 
The dashed horizontal lines show the cross-sections, for which we present conductance curves (Figs. \ref{fig2}, \ref{fig3}, \ref{fig4}).
(b) Partial conductance of TQD as a function of $\lambda_{I}$ for several values of Fano factor $(E_{f}=-1.5)$, inset shows occupation of IQD. 
(c), (d) Transmissions for $q=\pm 0.5$ and $\lambda_{I}=0, 0.2$ (c), $\lambda_{I}=0.5$ (d) and $\lambda_{I}=0.8$ (inset of Fig. d).}
\label{fig3}
\end{figure*}
Charge stability diagram of TQD as a function of IQD energy $E_{f}$ and e-ph coupling constant $\lambda_{I}$ 
is presented on Fig. \ref{fig3}a. It is seen that transfers between regions of different occupations are possible either by changing gate voltage (change of the dot single particle energy $E_{f}$) 
or by modification of electron-phonon coupling. Polaron induced suppression of charging energies shifts the Coulomb blockade boundaries and narrows Coulomb valleys. 
Possible transitions are: $n=1\rightarrow n=2,$ $n=0\rightarrow n=1\rightarrow n=2$ and $n=0\rightarrow n=2$. 
Let us first concentrate on phonon induced $1\rightarrow 2$ transition (Fig. \ref{fig3}a). 
For $n=1$ SU(2) Kondo resonance forms on IQD. Due to the interference with the wave propagating through the OQD Fano-Kondo resonance appears determined by the value of $q$, 
which in this case does not change with the strength of e-ph coupling $\lambda_{I}$. In consequence, Fano-Kondo conductances remain almost unchanged in the whole range of single occupation. 
For $q=0$ antiresonance blocks the linear transport. The asymmetry of the Fano line does not change with $\lambda_{I}$, 
but the width of the dip does due to phonon induced renormalization of interdot hopping $t’$ and polaron shifts of site energy and Coulomb interaction of IQD 
(compare transmissions for $\lambda_{I}=0$ and $\lambda_{I}=0.2$ (Fig. \ref{fig3}c)). Total Fano-Kondo conductance takes the value of zero for $q=0$ (antiresonance) and $e^{2}/h$ for $q=\pm 1$.
For $\lambda_{I}=0.5$ a transition $1\rightarrow 2$ to the new charge state takes place. The shapes of the transmission lines are combined effect of charge transition and interference. 
Obviously the widths of the dips appearing here are wider than for Fano-Kondo resonances. $T(E_{F})$ reaches one for $q<0$ and it takes the value zero for $q>0$. 
This is illustrated on Fig. \ref{fig3}d for the exemplary case of $q=\pm 0.5$, but these limits are valid for any value of $q$. 
This behavior reflects in conduction through its jump to the unitary limit or in its complete suppression in the transition point. 
For $\lambda_{I}>0.5$ occupancy increases to $n=2$ and the fully occupied IQD stops affecting transport. 
Transmission in this range does not result from interference and the observed dependence on $q$ only reflects dependence on the site energy of the open dot $E_{O}$. 
The widths of the lines are now determined by unperturbed dot-lead coupling (inset of Fig. \ref{fig3}d). 
Next picture (Fig. \ref{fig4}) presents direct transition in TQD from the empty into double occupied state $0\rightarrow 2$ for $q=0$ with phonon coupled to interacting dot. 
\begin{figure*}
\centering
\includegraphics[width=12cm]{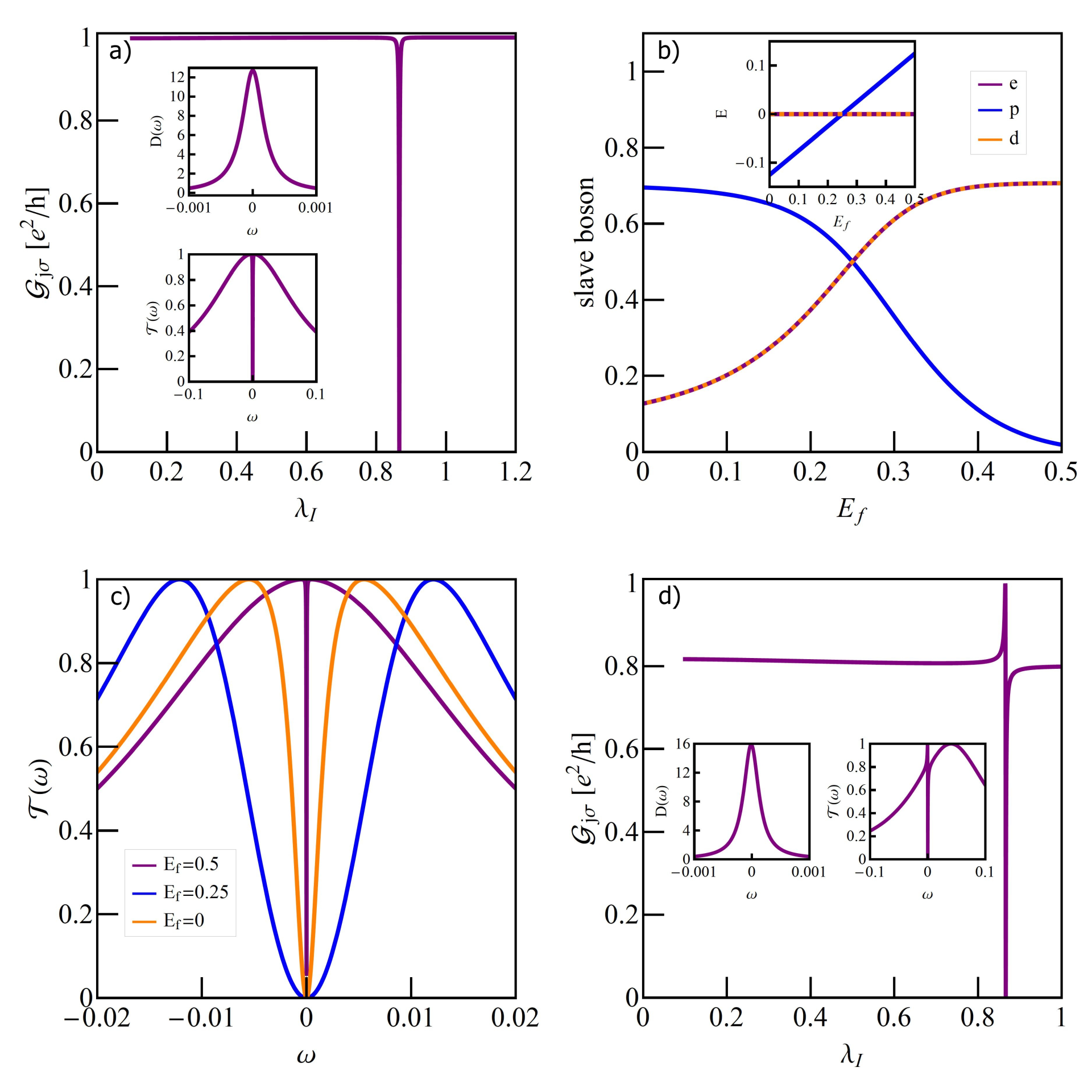}
\caption{(a) Partial conductance of TQD for $E_{f}=0.5$ with visible effect of charge Kondo resonance at $\lambda_{I}=0.43$. 
Top inset shows density of states of IQD and bottom one transmission in charge Kondo state (b) Gate dependence of slave bosons with $\lambda_{I}$ on the line of e-d degeneracy. 
(c) Several representative transmissions for different gate voltages (d) Partial conductance for $E_{f}=0.5$ and $q=0.5$ together with DOS of IQD and transmission at the Kondo point.}
\label{fig4}
\end{figure*}
Empty $(\lambda_{I}<0.43)$ or fully occupied IQD $(\lambda_{I}>0.43)$ does not affect transport.
The electron is transported through the OQD in these ranges with probability amplitude $T(E_{f})=1$. At $\lambda_{I}=0.43$  
the effective $0\rightarrow 2$ charge fluctuations (pseudospin fluctuations) lead to the formation of charge Kondo resonance on the interacting dot, which interfering with the wave propagating through an open dot 
causes the charge Fano-Kondo resonance to occur. This reflects in the occurrence of the dip of conductance. The resonance peak at the interacting dot and the corresponding symmetric dip of transmission are shown in the insets. 
They are considerably narrower than spin Fano-Kondo dips from Fig. \ref{fig3}c.
Figure \ref{fig4}b shows boson amplitudes for different dot energies $E_{f}$ and e-ph coupling $\lambda_{I}$ chosen such that empty and double occupied states degenerate. 
In charge Kondo state the low energy excitations are charge fluctuations and there is a large gap for spin fluctuations. In SB language it means that equal $e$ and $d$ amplitudes are close to the value $e=d\approx 1/\sqrt{2}$ 
corresponding to the transport unitary limit and $p_{\sigma}$ amplitudes are small. As it is seen this condition is fulfilled for $E_{f}=0.5$ for which dependencies from Fig. \ref{fig4}a are drawn. For lower values of $E_{f}$ 
the role of spin fluctuation increases as can be seen from the increase in $p_{\sigma}$ at the expense of $d$ and $e$ amplitudes. This fact is also demonstrated by the energy dependences of the states corresponding to $n=0, 1, 2$ 
drawn versus $E_{f}$ in the inset of Fig. \ref{fig4}b. This also reflects in broadening of the resonance line on the open dot as shown in Figure \ref{fig4}c. 
The presented transmissions correspond to charge Fano-Kondo state, mixed valence and spin Fano-Kondo state respectively. For $E_{f}=U/2$ all four states degenerate and below this value 
spin fluctuations gradually take over the leading role. In the region of low values of $e$ and $d$ amplitudes the spin Kondo resonance will form. This time the amplitudes of $p_{\sigma}$ operators take the values close to $1/\sqrt{2}$. 
Fig. \ref{fig4}d shows examples of perturbed charge Fano-Kondo effects for asymmetric cases $q=0.5$. As expected resonance lines on the IQD are wider in this cases. 
Since for $n=0$ and $n=2$ interacting dot is decoupled from transport path the presented conductance in these ranges is fully determined by position of $E_{O}$ i.e. by the value of $q$. 
Fig. \ref{fig5} presents gate dependence of conductance for the discussed cases. For $q\neq 0$ interference introduces asymmetry of conductance with respect to e-h symmetry point $E_{f}=-U/2$. 
Since phonon coupling with OQD changes interference conditions symmetric conductance for $q_{0}=0$ becomes asymmetric for $\lambda_{O}\neq 0$. On the contrary, coupling with IQD $(\lambda_{I}\neq 0)$ maintains symmetry. 
\begin{figure}
\centering
\includegraphics[width=6cm]{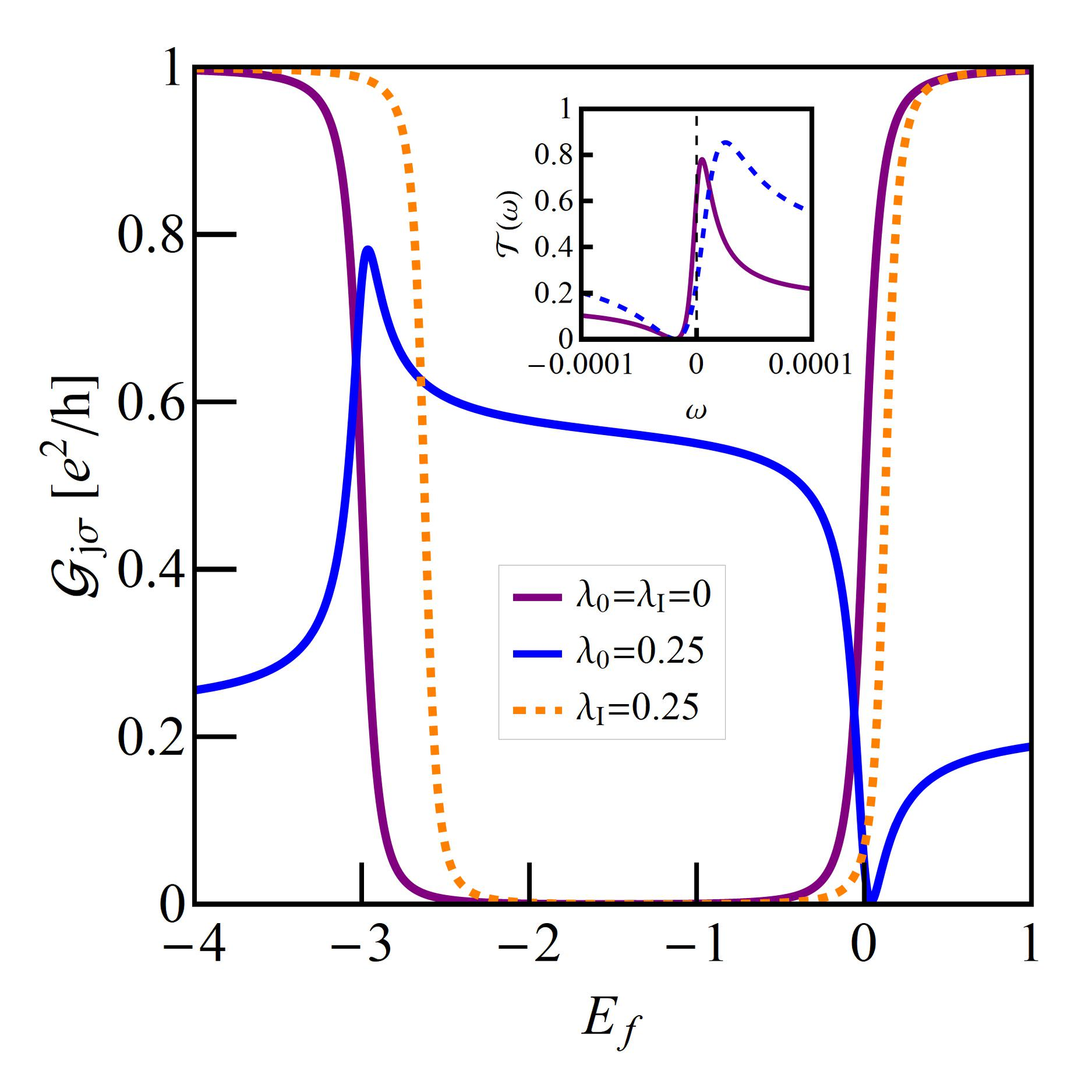}
\caption{a) Comparison of gate voltage dependencies of conductance of TQD with phonon coupled to IQD (symmetric orange broken line) or OQD (asymmetric blue line) with conductance of TQD decoupled from phonons. 
Inset presents transmission for $\lambda_{O}=0.25$ for $E_{f}= -0.2$ (dashed blue line) and $E_{f}= -2.8$ (solid purple line).}
\label{fig5}
\end{figure}
Fig. \ref{fig6} compares the dependencies of Kondo temperatures on the strength of e-ph coupling for the phonon mode coupled either to the open dot or to the interacting dot for TQD structure (SU(2) symmetry) and for DTQD with single phonon mode (SU(4)). 
The latter problem is discussed in the next section. It is seen that in both cases, the decrease of $T_{K}$ with the increase of $\lambda$ goes faster for the case of phonon attached to the open dot than, 
when phonon is connected to the interacting dot. This indicates that among the polaron effects, the exponential decay of the hopping integrals has the dominant impact on the lowering of Kondo temperature. 
Before discussing the results for DTQD let us first comment on the SU(4) symmetric case to which we refer in the analysis below. 
As already mentioned in the introduction, many experimental facts \cite{ref10, ref13, ref14, ref15, ref16, ref17} are attributed to the observation of this symmetry. 
Although the fully symmetric double dot system is only an idealization, some studies suggested \cite{new15}, 
that emergent low energy SU(4) symmetry can be restored also for slightly asymmetric systems by appropriate adjusting the gate voltages. 
More recent analysis based on NRG calculations \cite{new16} 
showed however, that for $U’<U$ the restoration of symmetry in the low energy range might only happen if the interdot interaction is greater than the half bandwidth of the leads, 
what is experimentally unrealistic. In our considerations, the equality of all pure Coulomb interactions is not crucial, because by changing the electron–phonon coupling they differentiate or equalize anyway. 
We assume equal values of $U$ and $U’$ only for clarity of presentation. Certainly the problem, whether the emergent SU(4) state could be reached through the e-ph coupling is interesting in itself, but this would require a more detailed analysis. 
\begin{figure}
\centering
\includegraphics[width=6cm]{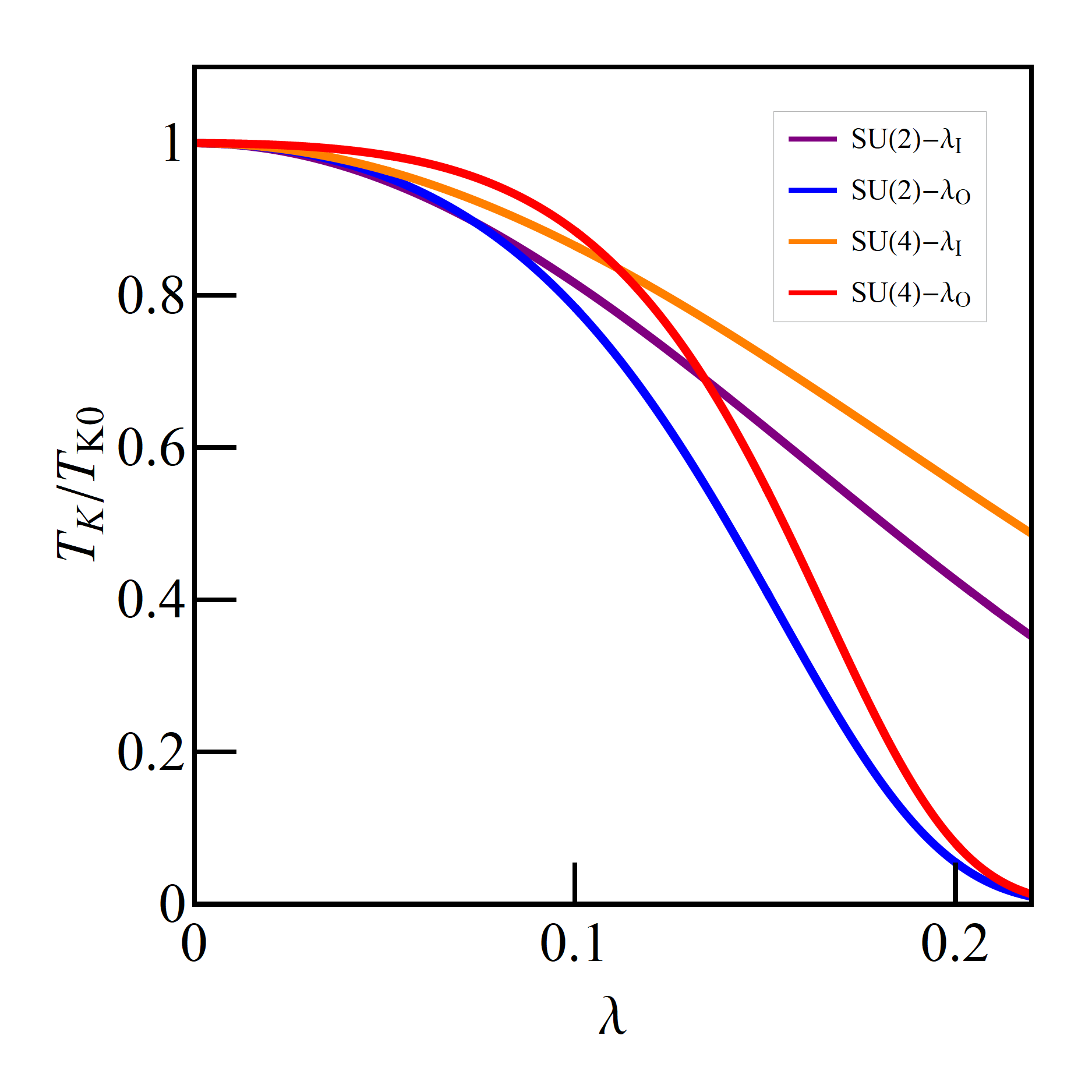}
\caption{Relative Kondo temperatures vs. e-ph coupling strength of TQD and DTQD coupled to the single phonon mode ($E_{f}=-1.5$).}
\label{fig6}
\end{figure}

\subsection{Double T-shaped quantum dot structure}

\begin{figure}
\centering
\includegraphics[width=6cm]{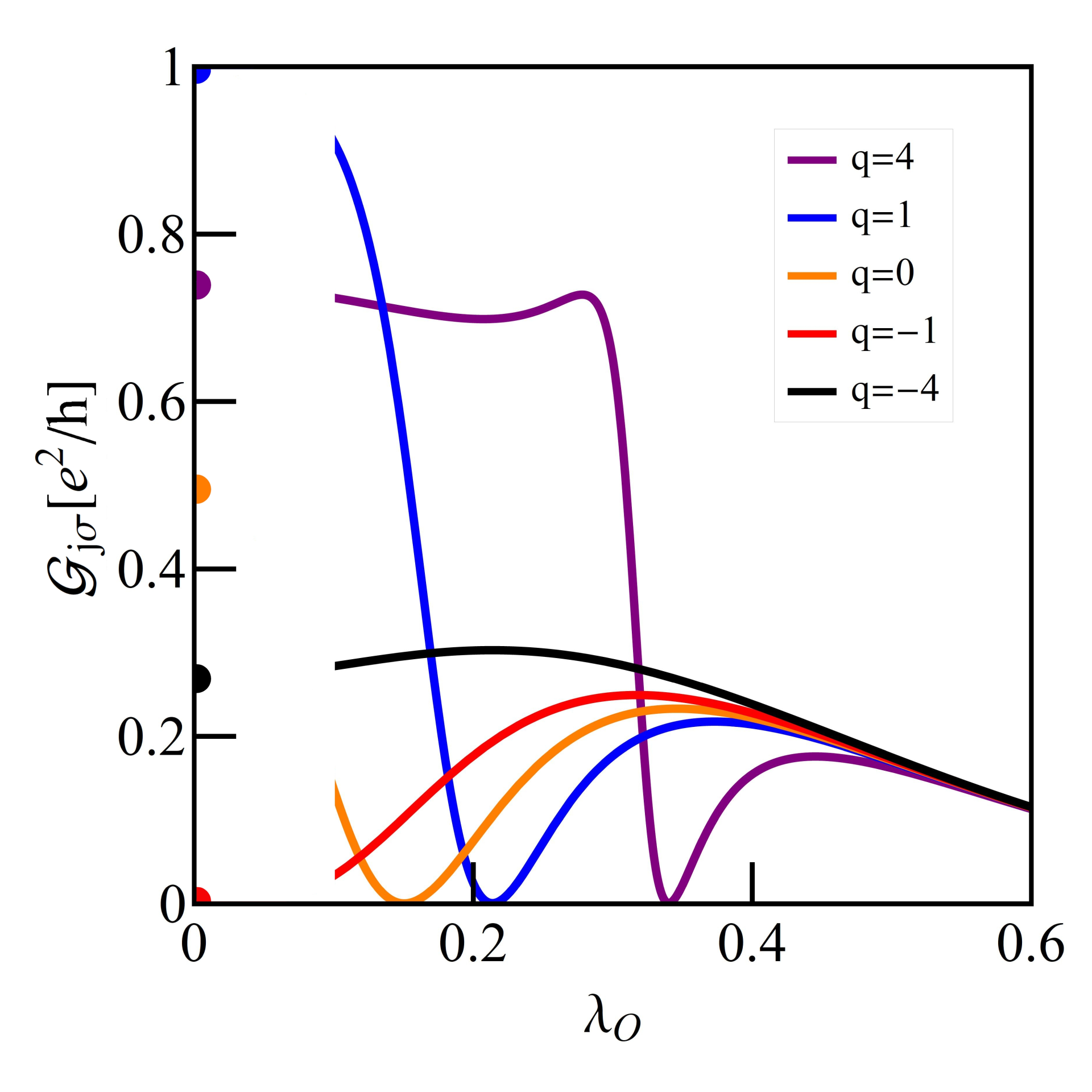}
\caption{Partial conductance of DTQD with phonons coupled to OQDs as a function of $\lambda_{O}$ for different Fano factors for $E_{f}= -1.5$.}
\label{fig7}
\end{figure}

\begin{figure*}
\centering
\includegraphics[width=12cm]{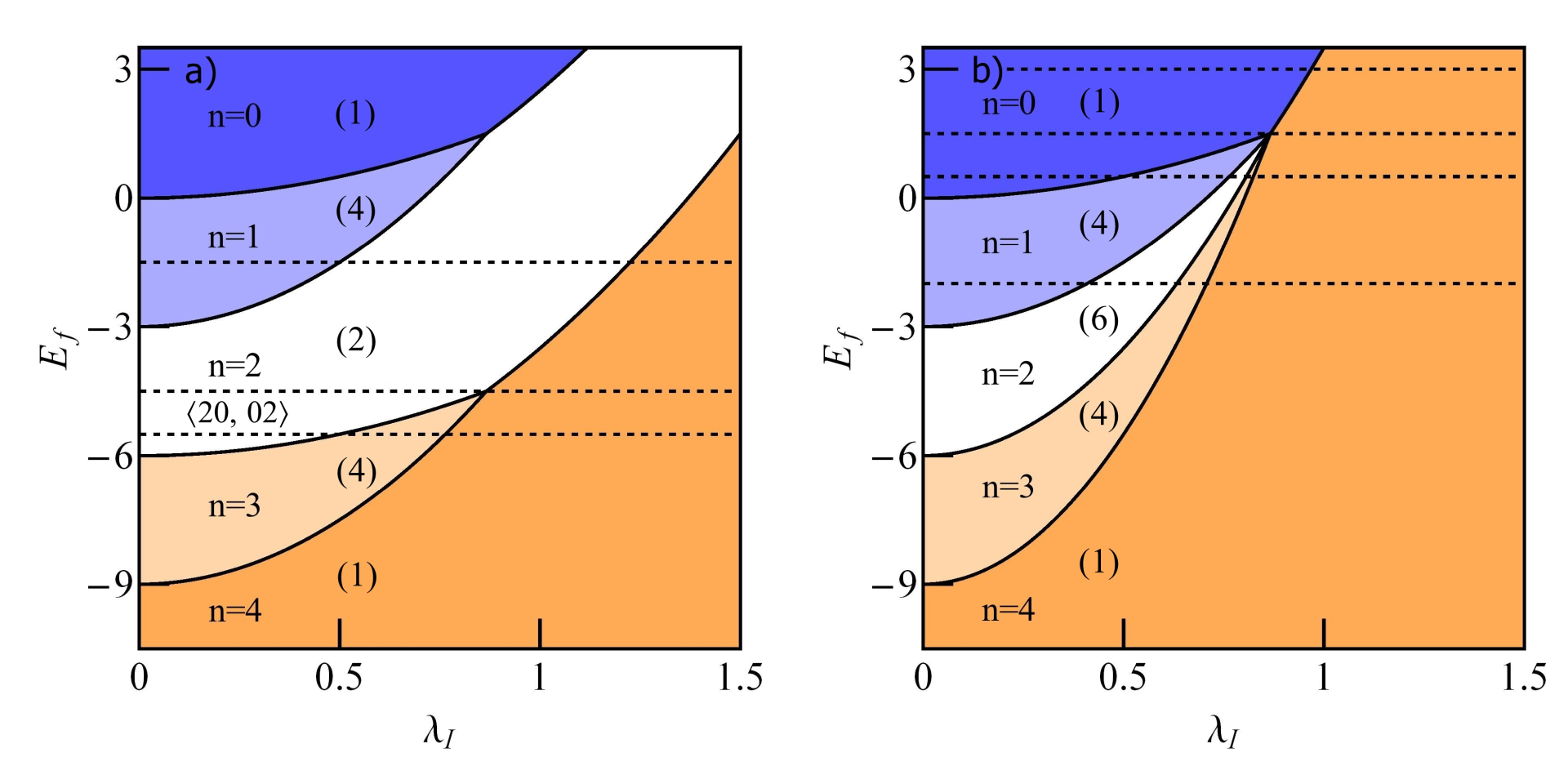}
\caption{(a) Charge stability diagram of DTQD with phonons coupled to the interacting dots as a function of gate voltage and e-ph coupling $\lambda_{I}$. 
Apart from occupation numbers also the corresponding degeneracies of the ground states are given in the brackets. The dashed horizontal lines indicate the cross-sections for which we present conductances below.
(b) Charge stability diagram of DTQD with single phonon coupled to both interacting dots.}
\label{fig8}
\end{figure*}

\begin{figure*}
\centering
\includegraphics[width=12cm]{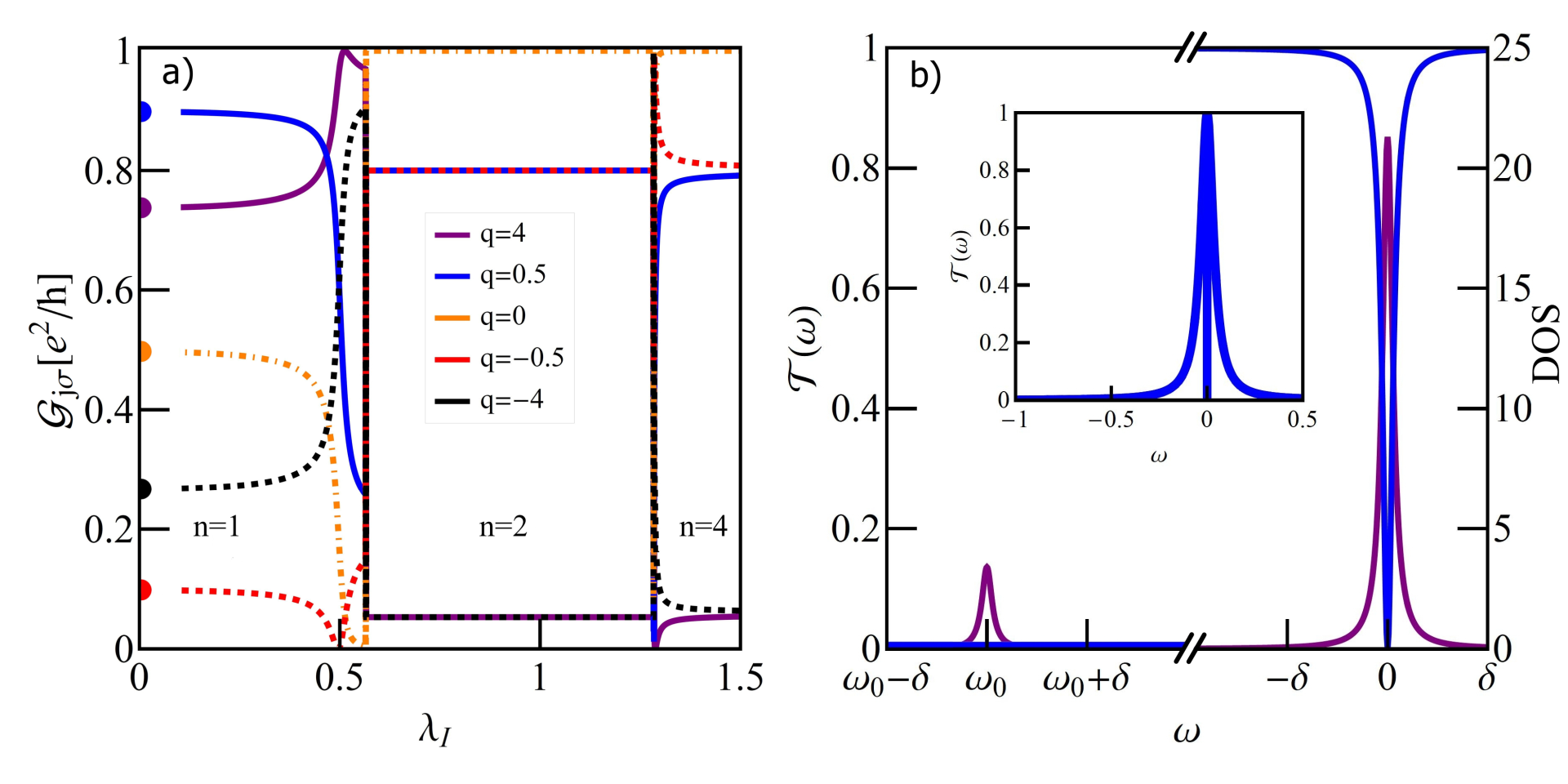}
\caption{(a) Partial conductances of DTQD with a pair of phonons as the functions of e-ph coupling $\lambda_{I}$ for $E_{f}=-1.5$, for different Fano factors 
(b) Density of states and transmission for the DTQD system with e-ph coupling $\lambda_{I}=0.2$, $E_{f}=-1.5$ and $\delta=0.001$. Phonon modes reflect in DOS by the appearance of satellites, but no traces of them are visible in transmission.
The full shape of transmission is shown in the inset.}
\label{fig9}
\end{figure*}

Fig. \ref{fig7} presents the impact of phonons coupled to the open dots on the conductance of DTQD system. 
For $\lambda_{O}=0$ Fano-Kondo resonance for $q=0$ reveals through half reflection, for $q=-1$ antiresonance is formed and conductance drops to zero and for $q=-1$ unitary conductance is observed. 
The observed dependencies are the combined effects of phonon induced renormalization of Fano factors $q_{eff}$ (change of the interference conditions) and Frank-Condon suppression. 
As mentioned earlier, the latter factor plays crucial role for large values of e-ph coupling, where all the conductance curves for different $q_{0}$ start to converge. 
The zeros of conductances are achieved when $q_{eff}$ reaches -1 (destructive interference). For $q_{eff}=1$ the phenomenon of constructive interference occurs (unitary transmission) and for $q_{eff}=0$ half reflection is observed. 
Fig. \ref{fig8} presents charge stability maps of DTQD when two phonons are attached to the interacting dots (Fig. \ref{fig8}a) and when a single phonon is coupled to both IQDs (Fig. \ref{fig8}b). 
The map is plotted versus interacting dot's energy $E_{f}$ and e–ph interaction constant $\lambda_{I}$. 
The degenerations of the ground states are also marked on the map. Two triple charge points are visible on map (a) ($n = 0, 1, 2$ and $n = 2, 3, 4$) i.e. points of coexistence of states characterized by three different charges. 
In each of these points degenerate seven states. Horizontal dashed lines indicate the cross-sections for which the plots of conductances are presented below. 
Interesting feature of the map is the occurrence of the quadruple charge point and the possibility of phonon induced $0\rightarrow 4$ charge transition. 
In quadruple charge point all sixteen double-dot states are degenerate.
Fig. \ref{fig9}a shows conductance vs. $\lambda_{I}$ for $E_{f}=-1.5$ 
(charge transition path $n=1\rightarrow n=2\rightarrow n=4$). 
For $\lambda_{I}=0$ DTQD exhibits SU(4) symmetry and interference between the paths through IQDs, where Kondo resonance is formed, 
with direct paths through OQDS leads to the appearance of Fano-Kondo resonance with conductance determined by Fano parameter q. For $q=0$ half reflection is observed for $\lambda_{I}=0$. 
E-ph coupling on the interacting dots does not change the interference conditions, it influences the correlations through the renormalization of the open dot – interacting dot coupling and shifts of the on-site energies of IQDs. 
Let us examine conductance starting from the unperturbed IQD level. E-ph coupling breaks SU(4) symmetry, because due to the coupling with phonons the intradot Coulomb interaction $U$ renormalizes, 
whereas interdot (or interorbital) interaction $U’$ does not change $(U\neq U’)$. This means that for $n=1$ spin-charge Fano-Kondo resonance is partially destroyed, which is manifested in the changes of conductance 
with the increase of $\lambda_{I}$. 
Fig. \ref{fig9}b compares density of states on the IQDs and corresponding transmission of OQDs dot for $\lambda_{I}=0.2$ $(n=1)$. 
This picture is included as a representative example illustrating the general feature of coupling of phonons to the electrons on the interacting dots. 
The satellite peaks appear only in DOS of IQD, but their traces are not visible in transmission. This is in contrast to that, what is observed in coupling to an open dot. 
Around $\lambda_{I}=0.5$ phonons induce transition to the new charge state $n=2$ (see map 8a) and in the transition region most drastic changes of conductances are observed. 
Beyond this region conductances become fixed and as it is easy to check their values correspond to transmission of the open dots determined only by $q$, 
which indicates that the interacting dots are disconnected from transport. There are six states characterized by occupation $n=2$, 
two with double occupancy at the single dot and vacant on the other dot ${(0,2), (2,0)}$ and four states with single occupancy on each of the dots 
${(\uparrow,\uparrow), (\uparrow,\downarrow), (\downarrow,\downarrow), (\downarrow,\downarrow)}$. 
For the discussed case of $\widetilde{U}<U’$ these first states are energetically lower by $U’ - \widetilde{U}$ than the latter. 
It is well know that for strong interdot interactions $(U’>>U)$ electrons prefer occupy the same dot i.e. 
the states $(2, 0)$ and $(0,2)$ dominate and the charge ordered state (CO) with fully occupied one of the dots becomes the ground state in $n=2$ region. 
For $U\rightarrow \infty$ the dots decouple from the leads, leaving a pair of free conduction bands and a dot with degenerate charge configurations (0, 2) and (2, 0) \cite{new22}.
It has been also predicted by the numerical renormalization group analysis performed for the system of capacitively coupled dots 
that CO state occurs already for small deviations of the values of intra and interdot interactions $(U’-U\geq T_{K}^{SU(4)})$ \cite{new23}. Similarly, one can also expect the same CO ground state with $(2,0)$ or $(0,2)$ states for DTQD. 
SBMFA formalism we use applied to model (2, 12), with no symmetry breaking perturbation does not point directly to CO as the ground state. 
Instead we found solution with two bosonic amplitudes $d1=d2=1/\sqrt{2}$ and exactly vanishing rest of the amplitudes. The found SB renormailzation factor $z$ (12) is equal zero, which indicates vanishing of the Kondo scale i.e. 
destruction of Kondo correlation at any finite temperature and decoupling of the interacting dots from the transport path. The obtained solution signals the destruction of the strongly correlated ground state. 
This suggests a conjecture, that needs however to be verified, that similarly to the isolated DTQD system, the ground state in this range is either in $(0, 2)$ or $(2,0)$ with probability $1/2$ 
and infinite time of transition from one degenerate state into another $(z=0)$. 
We have obtained clear evidence of charge ordered ground state in $n=2$ domain within SBMFA formalism by supplementing Hamiltonian (2) by a small symmetry breaking term. 
Similarly, as has been proposed in \cite{ref72}, we have introduced potential scattering correlated to dot occupancy $H_{K}=K\sum_{j\sigma}(n_{j}-1)c_{0j\sigma}^{\dagger}c_{0j\sigma}$ 
This perturbation stabilizes symmetry-broken CO state $(d1\approx 1, d2\approx 0$ or $d2\approx 1, d1\approx 0)$. 
Hereafter this approach is called by us SB1 and is used by us only in $n=2$ region.
\begin{figure*}
\centering
\includegraphics[width=12cm]{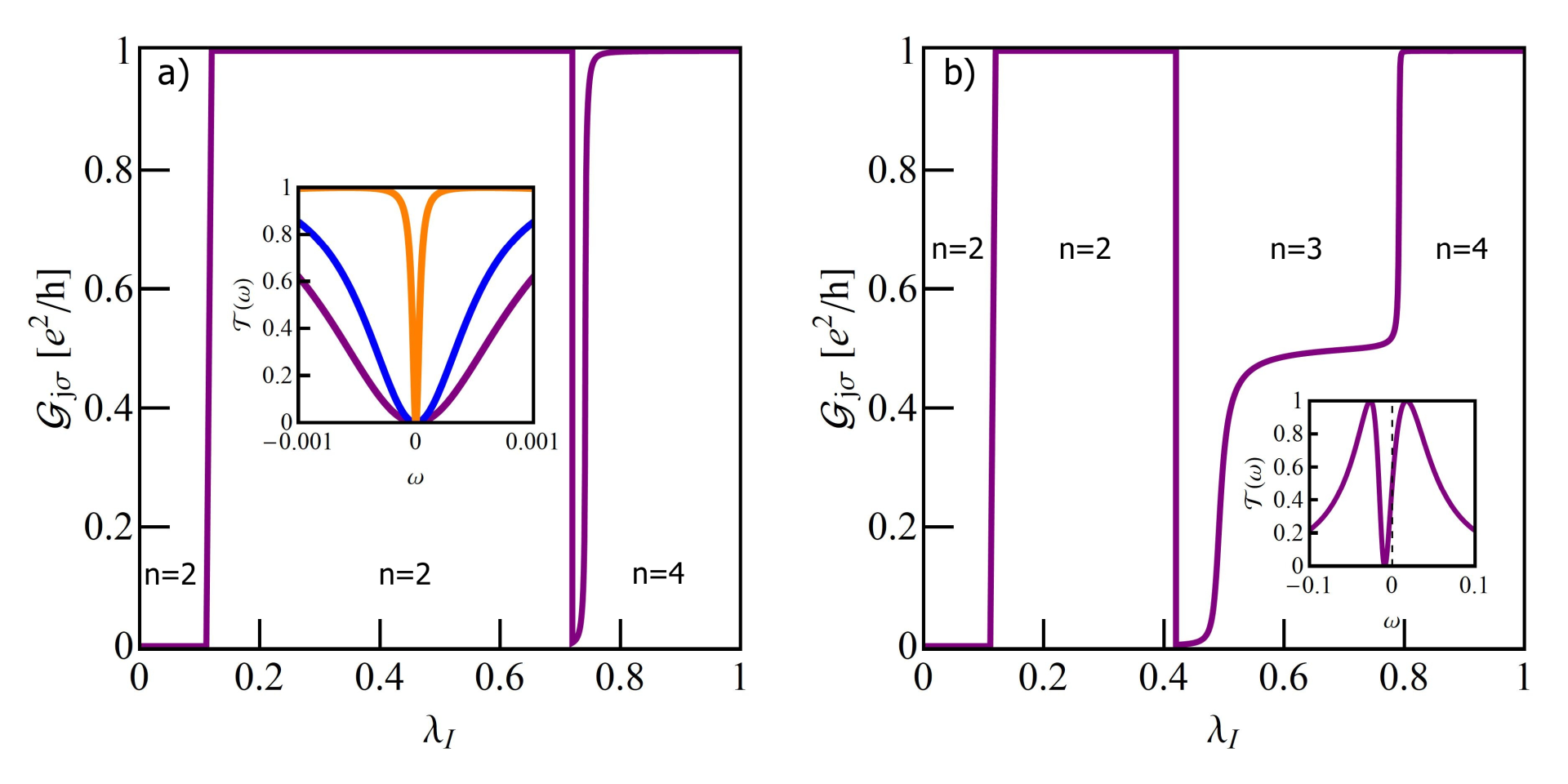}
\caption{(a) Partial conductance of DTQD with a pair of phonons as a function of e-ph coupling $\lambda_{I}$ with $E_{f}=-4.5$. For $\lambda_{I}<0.1$ system is in the broken spin-orbital Kondo state, 
for $\lambda_{I}=0.1$ charge Kondo state emerges and for $\lambda_{I}>0.1$ system enters into charge ordered state $(2, 0)$ or $(0, 2)$. 
Around $\lambda_{I}=0.72$ transition to the fully occupied state is observed. Inset shows transmissions for $\lambda_{I}=$0 (purple), 0.085 (blue) and 0.095 (orange) 
(b) Analogous conductance dependence as in (a), but for $E_{f}=-5.5$ with additional transition to spin Kondo state for $n=3$. Inset present transmission for $\lambda_{I}=0.65$}
\label{fig10}
\end{figure*}
Now let us look at the conductance for $q=0$, but starting from dot energy $E_{f}=-4.5$, which already for vanishing e-ph coupling corresponds to double occupancy $(n=2)$ (Fig. \ref{fig10}). 
In this case for $\lambda_{I}=0$ SU(4) Kondo resonance is formed in the interacting dots, where all six degenerate states corresponding to $n=2$ are engaged in spin-charge fluctuations. 
The resulting Kondo resonance is centered at $E_{F}$ and correspondingly SU(4) Fano-Kondo antiresonance has the shape of symmetric dip with zero value at the Fermi level (see inset of Fig. \ref{fig10}a). 
Zero transmission corresponds to zero conductance. For increasing coupling the symmetry is gradually broken and antiresonance narrows down (transmission for $\lambda_{I}=0.085$). 
Withe the increase of $\lambda_{I}$ $U'$ effectively increases with respect to $U$ making the charge states ${(0,2), (2,0)}$ 
lower in energy by $U’-U=2\lambda_{I}^{2}/\omega_{0}$ than the spin states ${(\uparrow,\uparrow), (\uparrow,\downarrow), (\downarrow,\uparrow), (\downarrow,\downarrow)}$. 
This leads to the transition to CO state for $\lambda_{I}>\lambda_{c}$($\lambda_{c}=0.1$, $2\lambda_{c}^{2}/\omega_{0} > T_{K}^{SU(4)}$), which is well reproduced by SB1 formalism. 
Before this transition happens charge Fano-Kondo state with fluctuating states ${(0,2), (2,0)}$ is formed. 
Inset of Fig. \ref{fig10} shows densities of states at the interacting dots for $\lambda_{I}<\lambda_{c}$. The considerable decrease of the widths of the resonance peaks is observed. 
For $\lambda_{I}=\lambda_{c}$, where charge pseudospin is quenched Kondo temperature becomes extremely small. 
The cotunneling processes are not direct, they are mediated by higher energy spin states $(1,1)$ and this results in the low Kondo temperature. 
Up to the coupling value $\lambda_{I}=0.72$ DTQD remains in the degenerate charge ordered states $(0,2)$ or $(2,0)$, and similarly as in the case discussed previously transport is not influenced by the interacting dots 
(full unitary transmission for $q=0$). The dip of conductance occurring around $\lambda_{I}=0.72$ reflects transition to the state with both dots fully occupied $2\rightarrow 4$ and for the discussed case of $q=0$ 
unitary conductance is observed for $n=4$. As it is seen transition region in the presently discussed case is much broader than presented on Fig. \ref{fig9} and it is related to the proximity on the energy scale of three-electron states 
$(n=3)$ (see map 8a). For slightly lower values of $E_{f}$, e.g. $E_{f}=-5.5$ before transition to the fully occupied state occurs, there will be a transition to $n=3$ first (Fig. \ref{fig10}). 
In the region of triple occupancy (single hole) spin-orbital SU(4) Fano-Kondo resonance is formed and half reflection is observed. 
The observed plateau of conductance $\mathcal{G}_{j\sigma}=0.5$ corresponds to the spin-orbital (spin-charge) SU(4) hole Fano-Kondo effect, in which participate states $\{|\uparrow \downarrow ,\sigma\rangle,|\sigma',\uparrow\downarrow\rangle\}$. 
Resonance transmission is drawn in inset of Fig. \ref{fig11}b. Next pictures concern the case when the single phonon mode is equally coupled to both interacting dots. 
In this case e-ph coupling does not break the full SU(4) symmetry.
\begin{figure*}
\centering
\includegraphics[width=12cm]{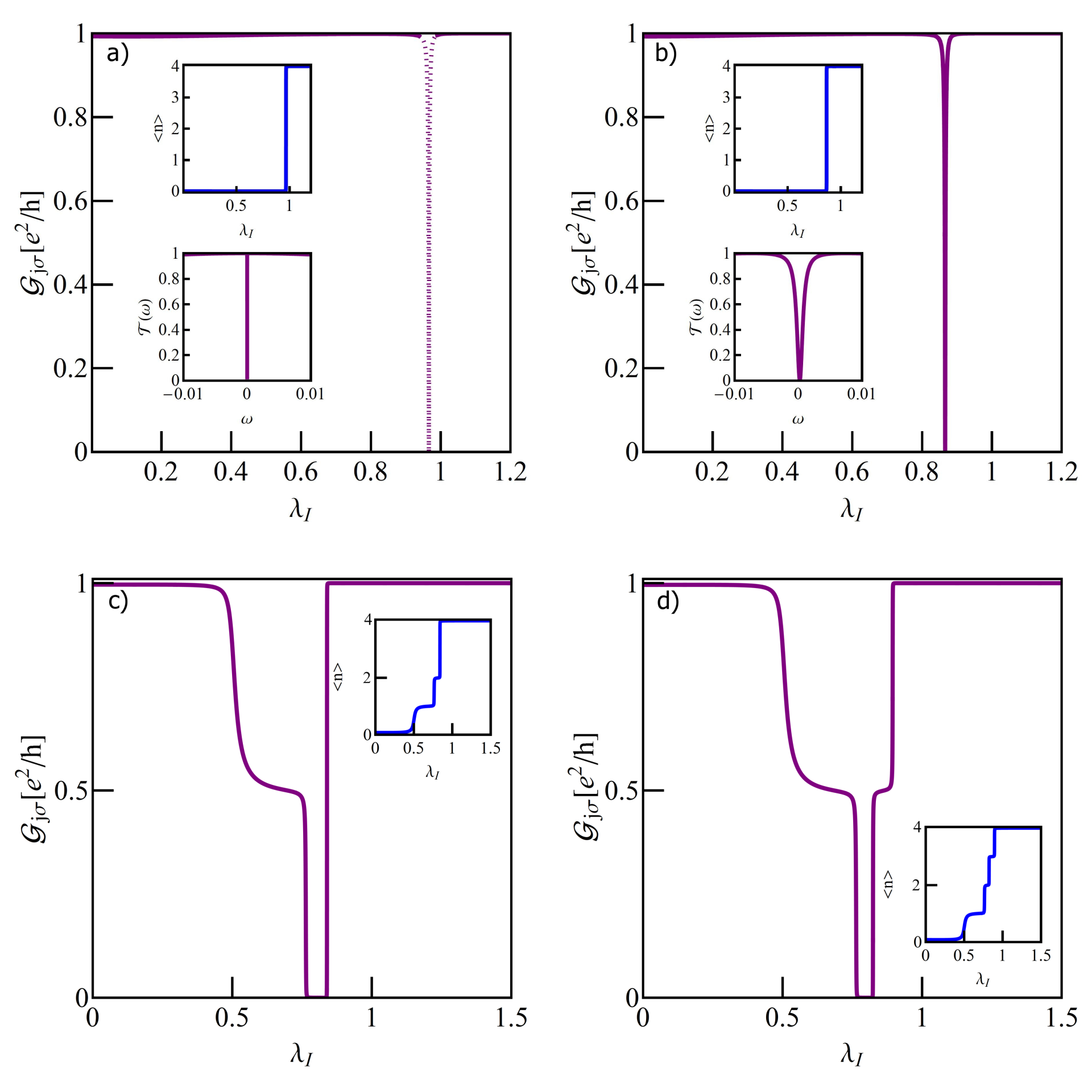}
\caption{Partial conductance and occupancies (insets) of DTQD coupled to a single phonon (a) $E_{f}=3$ ($0\rightarrow 4$ transition with charge Kondo state for $\lambda_{I}\approx 0.96$), 
(b) $E_{f}=1.5$ ($0\rightarrow 4$ mixed valence type transition), (c) $E_{f}=0.5$ ($0\rightarrow 1\rightarrow 2\rightarrow 4$ sequence of transitions) 
(d) $E_{f}=-2$ ($0\rightarrow 1\rightarrow 2\rightarrow 3\rightarrow 4$ sequence of transitions). 
Additional lower insets of (a) and (b) show corresponding transmissions close to the charge Kondo state ($\lambda_{I}=0.96$ (a)) or in charge transition point ($\lambda_{I}=0.85$ (b)).}
\label{fig11}
\end{figure*}
Fig. \ref{fig11} shows, apart from the exemplary conductances drawn for $q=0$ and chosen energies, also the corresponding dependencies of occupations on the strength of e-ph coupling (insets).  
For $E_{f}=3$ and $E_{f}=1.5$ direct transition $0\rightarrow 4$ is observed with no intermediate states. The interacting dots are disconnected from the transport when they are empty $(n = 0)$ 
or when the two dots are fully occupied $(n=4)$. In these cases the full transmission through the OQDs occurs giving total conductance $G_{tot}=2 e^{2}/h$. 
However in the transition point, the two mentioned cases are clearly different. 
For $E_{f}=1.5$ all sixteen states of different occupancies are degenerate in the transition point $(e=p_{j \sigma}=d_{j}=d_{\sigma\sigma'}=t_{j \sigma}=f=1/4)$. 
Interacting dots in this case are in mixed valence state and the real charge fluctuations between different fillings take place. The corresponding low energy transmissions at the open dots exhibit dips with 
widths only slightly reduced compared to the case of DTQD decoupled from phonons $(z=1)$. For $E_{f}=3$ two states degenerate at $\lambda_{I}=\lambda_{CK}$: empty and fully occupied, 
and the rest of the states are energetically distant. Low temperature physics of DTQD in this case plays out between these two degenerate states. Effective charge fluctuations between the empty dot system 
and quadruple occupied state can be considered as isospin flips of "up" $(n =0)$ and "down" $(n=4)$ states $(e=f=1/\sqrt{2})$. 
Flips correspond to a coherent movement of four electrons into and out of the system of coupled interacting dots in DTQD. 
One can suspect the occurrence in this point of the novel charge Kondo effect for $\lambda_{CK}=0.9604$ and the interference with direct paths leads then to the charge SU(2) Fano-Kondo antiresonance.  
Unfortunately, probably due to the extremely low Kondo temperature we have not succeeded numerically to find the SBMFA solutions for $E_{f}=3$ at the very point of transition $(\lambda_{I}=\lambda_{CK})$. 
Therefore we present in the inset of Fig. \ref{fig11}a transmission only in vicinity of $\lambda_{CK}(\lambda_{I}=0.96)$. The corresponding characteristic temperature in this point is $T_{K}\approx 10^{-9}$. 
Due to numerical uncertainty in the immediate vicinity of $\lambda_{CK}$ the conductance and occupation lines presented on Fig. \ref{fig11}a are drawn by the dashed lines. 
Transition to this charge Kondo state together with the examination of its stability under symmetry breaking perturbations is left for the future more elaborated study. 
Fig. \ref{fig11} illustrates also two other cases, where between occupations 0 and 4, intermediate fillings occur for in-between values of coupling strengths. 
Fig. \ref{fig11}c shows conductance for phonon induced $0\rightarrow 1\rightarrow 2\rightarrow 4$ transition and 
Fig. \ref{fig11}d for $0\rightarrow 1\rightarrow 2\rightarrow 3\rightarrow 4$. The observed intermediate plateaus for $n=1,2$ and in the latter case also for $n=3$ reflect the spin-charge (spin-orbital) SU(4) Fano-Kondo effects. 
The differences in the heights of plateaus for odd and even occupations are visible. In the former case the Kondo peaks at the interacting dots are shifted from the Fermi level and half reflection occurs at OQDs. 
In the range $n=2$ six two electron states are involved in cotunneling processes and in this case Kondo resonances are centered at $E_{F}$, what results in total suppression of conductance (SU(4) Fano–Kondo antiresonance).  

\section{Conclusions}

Our analysis of the interplay of interference, strong correlations and electron-phonon coupling is mainly addressed to molecular systems, where a strong coupling of local vibrations with electrons is expected. 
The discussion is also suitable for systems of suspended semiconductor-based quantum dots. 
It was found that transport through electron-phonon cavities with QDs embedded in a freestanding membrane is strongly affected by vibrational degrees of freedom.
As the model systems we have chosen single (TQD) or double (DTQD) T-shaped arrangements of quantum dots. The first system in the absence of coupling with phonons is characterized by SU(2) symmetry and the second by SU(4). 
An equivalent set to DTQD is single TQD with orbital degeneracy. To get a more complete insight into the issues discussed we considered local phonon modes coupled either to open, 
noninteracting dots connected directly to the leads or vibrations coupled to the interacting dots linked to the electrodes indirectly via the open dot. 
Phonons interacting with electrons in the open dots form polaron and effectively renormalize coupling to the leads and shift dot site energies. This changes the interference conditions and partially 
suppresses correlations on the interacting dot. In consequence modification of Fano-Kondo transmission is observed. The phononic effect gives also rise to the Franck-Condon suppression of conductance. 
Apart from the low energy resonance dip, transmission exhibits also satellite dips located at energies, which coincide with multiples of phonon energy. 
Due to the interactions of phonons with electrons on the interacting dot in turn not only dot level position, but also electron-electron interaction undergo a polaronic shift affecting the correlations. 
This type of coupling does not change interference conditions. Renormalizations of dots electron parameters also introduce modifications of charge stability diagrams. 
The distance between the boundaries of adjacent regions changes with the value of the e-ph coupling parameter. In different occupation regions apart from spin and spin-orbital Fano-Kondo resonances also 
charge Fano-Kondo effects are expected. The latter occur when phonons induce the effective attraction between electrons and states with even occupancy become energetically favored. 
When gate voltage is properly adjusted the energies of empty state and double or fourfold occupied states become degenerate. The low temperature dynamics is then described only by charge fluctuations – flips of charge pseudospin 
(charge Kondo effect). Of special interest are the effective fluctuations between degenerate empty and fully occupied states occurring in TQD or DTQD coupled to a single phonon mode 
(in the latter case phonon is equally coupled to both interacting dots (orbitals)). Double T-shaped system with two phonons separately coupled to different interacting dots is characterized by two different Coulomb interaction 
parameters, only one of which (intradot interaction) is renormalized by e-ph coupling. Phonon induced transition from single to double total occupancy leads for $n=2$ to charge ordered state on the interacting dots 
(either $(2,0)$ or $(0,2)$), which results in full transmission through the open dots. For gate voltages for which DTQD in the absence of phonons is already doubly occupied the interaction with phonons leads with 
increasing coupling parameter to the evolution from SU(4) symmetry with six degenerate states to broken SU(4), where the degenerate pair $(2,0)$ and $(0,2)$ of charge pseudospin separate from spin states. 
The charge doublet lies lower on the energy scale than spin quartet. When phonon induced decrease of interdot interaction parameter exceeds energy of SU(4) Kondo resonance a transition to charge ordered state occurs 
preceded at the transition point by the occurrence of charge Kondo effect with quenched charge pseudospin. 

Although T-shaped quantum dot systems discussed by us are only toy models, with the help of which we analyze the richness of emerging phenomena resulting from the interplay of three important factors: strong correlations, 
interference and coupling with phonons and discuss their impact on transport on a nanoscopic scale, the obtained results can be also qualitatively related to some experimental observations. 
Apart from many reports mentioned in the Introduction, which demonstrate phonon induced symmetric Kondo satellites, STM experiments presenting asymmetric satellite lines indicate the role of interference \cite{new17}. 
From the line widths of the central peaks or widths of satellite resonances one can infer about the Kondo temperature. In agreement with our calculations, it is observed, that $T_{K}$ increases with a weakening of e–ph coupling \cite{new24}. 

\bibliography{manuscript}

\end{document}